\documentclass[aps,prd,10pt,twocolumn,preprintnumbers,superscriptaddress,nofootinbib,letterpaper,longbibliography]{revtex4-2}

\usepackage{hyperref}
\usepackage{graphicx}
\usepackage{amsfonts,amsmath,amssymb,bm}
\usepackage{array}
\usepackage{xcolor}
\usepackage[utf8]{inputenc}
\usepackage{tikz}
\usepackage{multirow}
\usepackage{cleveref}
\crefname{equation}{Eq.}{Eqs.}
\crefname{figure}{Fig.}{Figs.}
\crefname{table}{Tab.}{Tabs.}
\crefname{section}{Sec.}{Secs.}
\crefname{appendix}{App.}{Apps.}
\Crefname{table}{Table}{Tables}
\Crefname{figure}{Figure}{Figures}
\hypersetup{
  colorlinks  = true,
  urlcolor    = blue,
  linkcolor   = blue,
  citecolor   = red
}

\newcommand{\ml}{m_{\ell}}

\newcommand{\costheta}{\cos\theta}

\newcommand{\El}{E_{\ell}}

\usepackage{orcidlink}

\newcolumntype{P}[1]{>{\centering\arraybackslash}p{#1}}

\def\Fermilab{Theory Division, Fermi National Accelerator Laboratory, Batavia, IL 60510, USA}

\graphicspath{{figures/}}

\begin{document}


\title{Machine Learning Neutrino-Nucleus Cross Sections II: More Leptons}

\author{Daniel C. Hackett \orcidlink{0000-0001-6039-3801}}
\email{danhack@uw.edu}
\affiliation{Institute for Nuclear Theory, University of Washington, Seattle, WA 98195-1550, USA}
\affiliation{\Fermilab} 

\author{Joshua Isaacson \orcidlink{0000-0001-6164-1707}}
\email{isaacs21@msu.edu}
\affiliation{Department of Physics and Astronomy, Michigan State University, East Lansing, MI 48824}

\author{Shirley~Weishi~Li~\orcidlink{0000-0002-2157-8982}}
\email{shirley.li@uci.edu}
\affiliation{Department of Physics and Astronomy, University of California, Irvine, CA 92697}

\author{Karla Tame-Narvaez\orcidlink{0000-0002-2249-9450}}
\email{karla@fnal.gov}
\affiliation{\Fermilab} 

\author{Michael L. Wagman \orcidlink{0000-0001-7670-1880}}
\email{mwagman@fnal.gov}
\affiliation{\Fermilab}


\begin{abstract}
Neutrino-nucleus cross sections are essential inputs for long-baseline neutrino oscillation experiments, yet remain challenging to model with high precision. Ref.~\cite{Hackett:2024dul} demonstrated that fully black-box data-driven cross-section models cannot accurately generalize between near- and far-detector fluxes, but imposing Standard-Model symmetries via structure-function parameterizations provides a well-posed learning problem in a simplified DUNE-inspired muon-neutrino disappearance analysis. Here, we extend this framework to multiple lepton flavors and both neutrino and antineutrino beam modes, incorporating the electron appearance channel and enabling a simultaneous determination of $\sin^2\theta_{23}$, $\Delta m^2_{31}$, and the CP-violating phase $\delta_{\rm CP}$. We find that the learned cross section model recovers the injected oscillation parameters without significant bias, with constraints comparable to those obtained using the exact underlying cross section. Flux uncertainties have more significant effects in the multi-channel case, and incorporating flux nuisance parameters into the learning problem is required for accurate results. Our results demonstrate that this data-driven approach can be extended to the multi-channel measurements needed for long-baseline neutrino oscillation analyses.
\end{abstract}

\preprint{FERMILAB-PUB-26-0703-T, UCI-HEP-TR-2026-14, MSUHEP-26-014, INT-PUB-26-039}

\maketitle


\section{Introduction}

Neutrino oscillation experiments are entering a precision era, and increasingly accurate theoretical inputs are needed to interpret the data~\cite{JUNO:2015zny,Hyper-Kamiokande:2018ofw,DUNE:2020ypp}. An important example is the neutrino-nucleus scattering cross section at GeV energies, which is critical for long-baseline accelerator neutrino experiments~\cite{Ruso:2022qes,deGouvea:2022gut,NuSTEC:2017hzk}. Accurate first-principles calculations of these cross sections remain challenging because accelerator neutrino energies lie in a regime where perturbative and nonperturbative QCD dynamics, as well as nuclear effects, all play a role~\cite{Andreopoulos:2009rq,Buss:2011mx,Golan:2012rfa,Hayato:2021heg,Isaacson:2022cwh,NuSTEC:2017hzk}.

Near detectors (NDs) provide high-statistics measurements of neutrino interactions before significant oscillations occur and play a central role in constraining cross section uncertainties. Conventionally, discrepancies between predicted and observed ND spectra are used to tune parameters of cross section models implemented in neutrino event generators before these models are applied to far-detector (FD) samples~\cite{NOvA:2020rbg,NOvA:2021nfi,T2K:2021xwb,T2K:2023smv}. The accuracy of this extrapolation, however, depends on how well the underlying interaction models describe the relevant physics~\cite{MINERvA:2019kfr,NOvA:2020rbg,Coyle:2022bwa,Coyle:2025xjk}.

In Ref.~\cite{Hackett:2024dul}, we explored a complementary, data-driven approach in which the neutrino-nucleus cross section is learned directly from ND data using machine learning (ML) techniques. Using Standard Model symmetries, we decompose the inclusive cross section into known kinematic factors and unknown nuclear structure functions. We represent these structure functions with neural networks (NNs), and train them directly on distributions of measured outgoing-lepton kinematics. This approach avoids event-by-event neutrino energy reconstruction and requires no assumed model for the underlying nuclear dynamics beyond the structure-function decomposition. It builds on successful demonstrations of NN parameterizations of one-dimensional parton distribution functions (PDFs) of nucleons~\cite{Ball:2011uy} and nuclei~\cite{AbdulKhalek:2019mzd} as well as two-dimensional structure functions in electron scattering~\cite{Forte:2002fg,Sobczyk:2024ajv,Bodek:2024mrp} and high-energy neutrino scattering~\cite{Candido:2023utz}. Importantly, this approach is not meant to replace event generator extractions of oscillation parameters, but to complement them.

We demonstrated the viability of this approach in Ref.~\cite{Hackett:2024dul} using a DUNE-inspired closure test. An NN cross section model was trained on simulated ND data generated with a known toy cross section and subsequently used to analyze simulated FD events. The oscillation results obtained using the learned cross section were comparable to those obtained in the idealized case where the underlying cross section was known exactly, including after incorporating several sources of systematic uncertainty. However, that initial proof-of-principle analysis considered only the muon-neutrino disappearance channel, $P(\nu_\mu\rightarrow\nu_\mu)$, and varied only $\sin^2\theta_{23}$ and $\Delta m_{31}^{2}$ in the oscillation analysis. In particular, electron appearance and antineutrino data, which are essential for measuring the CP-violating phase $\delta_{\rm CP}$, were left for future work.

Extending the method to multiple oscillation channels introduces several new ingredients. The cross section model must simultaneously describe electron and muon final states and both neutrino and antineutrino scattering. The inclusion of different charged-lepton masses provides additional information about the individual nuclear structure functions. At the same time, because argon is not an isoscalar target, the structure functions describing neutrino and antineutrino scattering are in general distinct, substantially enlarging the cross section inference problem. Importantly, accelerator neutrino experiments generally cannot distinguish electrons from positrons on an event-by-event basis, leading to contamination between the electron-neutrino and electron-antineutrino samples.

In this work, we extend the data-driven framework of Ref.~\cite{Hackett:2024dul} to this multi-channel setting. We consider both electron and muon final states and both neutrino and antineutrino beam modes, resulting in eight distinct event samples in our DUNE-inspired closure test. We train a common NN model for the full set of nuclear structure functions using the corresponding ND event distributions and apply the learned cross sections to a joint FD oscillation analysis. We simultaneously infer $\sin^2\theta_{23}$, $\Delta m_{31}^{2}$, and $\delta_{\rm CP}$, and compare the resulting constraints with the theoretical benchmark obtained when the exact underlying cross section is used. We find that the learned cross section model recovers the injected oscillation parameters without significant bias, with constraints comparable to those obtained using the exact cross section.

As in Ref.~\cite{Hackett:2024dul}, our goal is to test whether the cross section can be learned sufficiently well from ND measurements to enable a reliable oscillation analysis, rather than to construct a realistic model of neutrino-argon scattering. We therefore again employ a fully known toy cross section so that the quality of the learned model can be directly assessed. The larger set of interaction and oscillation channels considered here provides a more stringent closure test of the data-driven approach and demonstrates that it can be extended to the multi-channel measurements needed for long-baseline neutrino oscillation analyses.


\section{Methods}

\begin{figure*}[t]
    \centering
    \includegraphics[width=0.8\textwidth]{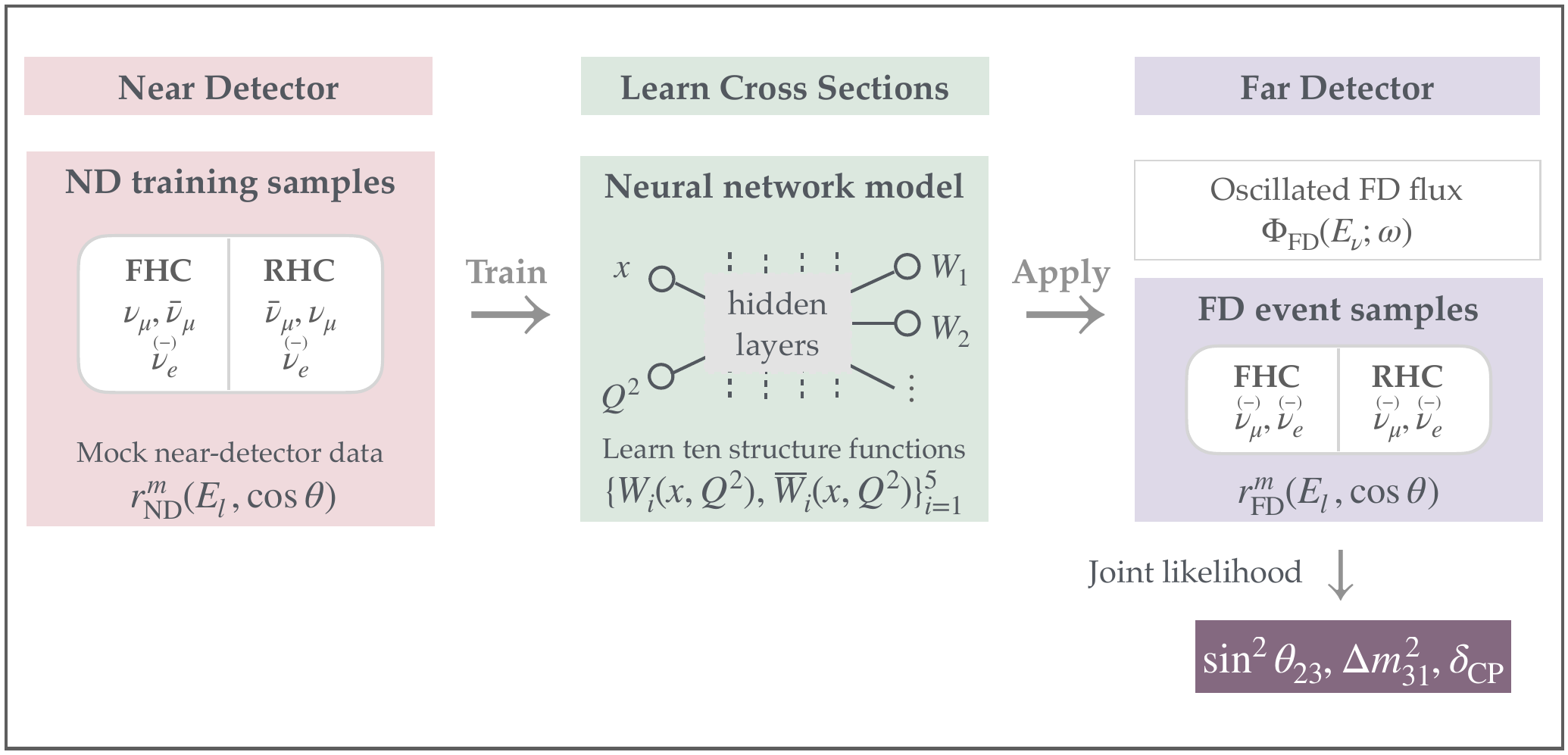}
    \caption{Schematic of the multi-channel data-driven oscillation-analysis framework. Near-detector event samples in both FHC and RHC modes are used to train neural-network representations of the neutrino and antineutrino structure functions. The learned cross sections are then combined with the oscillated far-detector fluxes to construct far-detector event samples and a joint likelihood for $\sin^2\theta_{23}$, $\Delta m^2_{31}$, and $\delta_{\rm CP}$.}
    \label{fig:analysis_workflow}
\end{figure*}

As mentioned above, the goal of this work is to use machine-learned cross sections for the extraction of oscillation parameters. In this section, we provide the details on the procedure. The full analysis strategy is summarized in Figure~\ref{fig:analysis_workflow}. First, the near-detector event samples in forward- and reverse-horn-current (FHC and RHC) modes are used to train neural-network representations of the neutrino and antineutrino structure functions. Second, the learned cross sections are then applied to the oscillated FD fluxes to construct FD event samples and a joint likelihood for $\sin^2\theta_{23}$, $\Delta m^2_{31}$, and $\delta_{\rm CP}$. We describe each step of this procedure in detail below.

\subsection{Cross sections and structure functions}

In analogy with the approach adopted in Ref.~\cite{Hackett:2024dul}, we parametrize the inclusive cross section in terms of structure functions (SF). However, in the present analysis, a more general treatment is required; specifically, a total of ten independent SFs are necessary to fully characterize the inclusive cross section for both neutrino and antineutrino scattering. 
The need for an increased number of SFs arises from the non-isoscalar nature of the argon target. 
Unlike an isoscalar nucleus, where the symmetry between proton and neutron contributions allows the SF parametrization for neutrino and antineutrino scattering to be equal, the proton-neutron asymmetry in argon implies that SFs must now be considered independently. 

The differential cross section for charged-current inclusive neutrino scattering can be expressed in terms of SFs as follows~\cite{Albright:1974ts,Paschos:2001np,Kretzer:2002fr}:
\begin{equation}
   \begin{split}
     &\frac{d^2 \sigma^{(\nu A)}}{d\El d\costheta}(E_\nu) = \frac{|V_{ud}|^2  G_F^2 \El}{\pi}   \sqrt{1 - \frac{\ml^2}{\El^2}}  \\
     &\hspace{25pt} \times \left\lbrace  \frac{E_{\nu}}{M_A} \left( 1 - y - \frac{Q^2 + \ml^2}{4 E_{\nu}^2} \right) W_2(x,Q^2)  \right. \\
     &\hspace{35pt} + \tilde{y} W_1(x,Q^2) + \left( 1 - \frac{\tilde{y}}{2} \right) W_3(x,Q^2)  \\
     &\hspace{35pt} \left.  - \left( \frac{\ml^2}{Q^2} \right) \left[ 2 W_5(x,Q^2) - \tilde{y}  W_4(x,Q^2) \right]  \vphantom{\frac{M_A xy}{2 E_{\nu}}} \right\rbrace,
   \end{split} 
   \label{eq:nuAd2}
\end{equation}
while the antineutrino version is
\begin{equation}
   \begin{split}
     &\frac{d^2 \sigma^{(\bar{\nu} A)}}{d\El d\costheta}(E_\nu) = \frac{|V_{ud}|^2  G_F^2 \El}{\pi}   \sqrt{1 - \frac{\ml^2}{\El^2}}  \\
     &\hspace{25pt} \times \left\lbrace  \frac{E_\nu}{M_A} \left( 1 - y - \frac{Q^2 + \ml^2}{4 E_\nu^2} \right) \overline{W}_2(x,Q^2)  \right. \\
     &\hspace{35pt} + \tilde{y} \overline{W}_1(x,Q^2) - \left( 1 - \frac{\tilde{y}}{2} \right) \overline{W}_3(x,Q^2)  \\
     &\hspace{35pt} \left.  - \left( \frac{\ml^2}{Q^2} \right) \left[ 2 \overline{W}_5(x,Q^2) - \tilde{y}  \overline{W}_4(x,Q^2) \right]  \vphantom{\frac{M_A xy}{2 E_{\nu}}} \right\rbrace .
   \end{split} 
   \label{eq:nuAd2anti}
\end{equation}
Here, $\theta$ denotes the scattering angle of the outgoing lepton, $m_\ell$ is the mass of the charged lepton, and $M_A$ represents the mass of the nuclear target. 
The squared four-momentum transfer is denoted by $Q^2$, while $x$ corresponds to the Bjorken scaling variable. 
The inelasticity is defined as $y = Q^2 / (2 M_A E_\nu x)$, and we introduce the modified inelasticity variable $\tilde{y} \equiv y(1 + m_\ell^2/Q^2)$ to account for finite lepton mass effects.

The nuclear SFs $W_i(x, Q^2)$ for neutrino interactions and $\overline{W}_i(x, Q^2)$ for antineutrino interactions are defined through the Lorentz-covariant decomposition of the forward matrix element $\langle A | J_\mu^\dagger J_\nu | A \rangle$, where $J_\mu = \bar{u} \gamma_\mu (1 - \gamma_5) d$ denotes the charged-current electroweak current, and $|A\rangle$ is the nuclear ground state.
We follow the convention of Ref.~\cite{Hackett:2024dul} and absorb factors of $x$ and $Q^2$ into the structure functions to remove kinematic zeros and poles from the coefficients. 
In terms of the functions $F_i$ of Ref.~\cite{Kretzer:2002fr}, our convention is
\[
W_i = x F_i \quad \text{for } i \in \{1, 3, 4, 5\}, \qquad W_2 = \left( \frac{2 x M_A^2}{Q^2} \right) F_2,
\]
with analogous definitions for the antineutrino functions $\overline{W}_i$. 
The terms involving $W_4$ and $W_5$ (or $\overline{W}_4$ and $\overline{W}_5$) are suppressed by $m_\ell^2/Q^2$. 
Their different contributions to electron and muon scattering provide additional information when the two final states are analyzed together.

Throughout this work, we neglect higher-order electroweak radiative corrections and terms of order $Q^2/m_W^2$. 
See Refs.~\cite{Tomalak:2021hec,Tomalak:2022xup,Afanasev:2023gev} for discussions of these effects.

The factors multiplying each structure function in Eqs.~\eqref{eq:nuAd2} and~\eqref{eq:nuAd2anti} define kinematic coefficients $K_{i}(E_\nu,E_\ell,\cos\theta)$ and $\overline{K}_i(E_\nu,E_\ell,\cos\theta)$ respectively, which allow the cross sections to be represented compactly as
\begin{equation}
    \frac{d^2 \sigma^{(\nu A)}}{d\El d\costheta}(E_\nu) = \sum_{i=1}^5 K_i(E_\nu,E_\ell,\cos\theta) W_i(x,Q^2),
\end{equation}
and 
\begin{equation}
    \frac{d^2 \sigma^{(\bar{\nu} A)}}{d\El d\costheta}(E_\nu) = \sum_{i=1}^5 \overline{K}_i(E_\nu,E_\ell,\cos\theta) \overline{W}_i(x,Q^2).
\end{equation}
This decomposition is central to our data-driven approach. 
The differential cross section depends on three independent kinematic variables, for example $(E_\nu,E_\ell,\cos\theta)$, whereas the ND measures a two-dimensional distribution in $(E_\ell,\cos\theta)$ after integration over the unobserved neutrino energy. 
A general three-dimensional cross section therefore cannot be inferred from that distribution alone. 
The SF decomposition supplies the needed constraint: Lorentz invariance fixes the kinematic coefficients $K_i$ and $\overline{K}_i$, while the unknown structure functions depend only on $(x,Q^2)$. 
We can thus fit these functions to flux-averaged ND data and use them with the oscillated FD flux, provided the ND data cover the $(x,Q^2)$ region relevant at the FD. 
For DUNE, the two detectors are expected to have substantial overlap in this region because oscillations primarily redistribute events within the same kinematic envelope~\cite{Hackett:2024dul}.


\subsection{Event rates and detector effects}
\label{subsec:detector}

For a given neutrino flux, the differential event rate with respect to the outgoing lepton energy $E_\ell$ and scattering angle $\theta$ is
\begin{equation}
\label{eq:rate}
    r^m_{D,\ell}(E_\ell,\cos\theta) = \int dE_\nu\, \frac{d^2\sigma}{dE_\ell\, d\cos\theta}(E_\nu) \, \Phi^m_{D,\ell}(E_\nu).
\end{equation}
Here, $m$ denotes the beam mode (FHC or RHC), $D$ labels the detector (ND or FD), $\ell$ indicates the charged lepton flavor (muon or electron), and $\Phi^m_{D,\ell}(E_\nu)$ is the corresponding neutrino or antineutrino flux.

In a realistic experiment, the measured lepton kinematics differ from their true values because of detector resolution. In a liquid argon time projection chamber (LArTPC), muons typically leave long tracks, while electrons and positrons produce electromagnetic showers, and the reconstruction quality for both muons and electrons is generally good. In this work, we approximate the charged lepton energy resolution by applying a $4\%$ Gaussian smearing to both electrons and muons~\cite{DUNE:2021tad}. There is also angular smearing for the outgoing leptons. The typical number quoted for DUNE is an angular resolution of $1^\circ$ for both electrons and muons~\cite{DUNE:2020ypp}. Because this is much smaller than our angular bin widths, we neglect angular smearing.

Another important detector effect to consider for our work is particle identification. 
Although most particles in a LArTPC can be identified through their characteristic energy-deposition patterns, important ambiguities remain. 
We account for the most relevant ambiguity in this work, charge identification, which determines which event rates can be analyzed separately. 
Electrons and positrons both produce electromagnetic showers and cannot be reliably distinguished event by event. 
For a beam containing both $\nu_e$ and $\bar{\nu}_e$, we therefore construct a single electron-like sample containing electrons from $\nu_e$ interactions and positrons from $\bar{\nu}_e$ interactions. 
At the ND, a magnetized muon tracker measures the charge of muons from their track curvature, allowing separate $\mu^-$ and $\mu^+$ samples for cross-section training. 
The FD has no magnetic field, so we combine the two muon charges into a single muon-like sample for the oscillation analysis. 
For either lepton flavor, the rate when charge is not identified is
\begin{equation}
\label{eq:tot_rate}
    r^{m}_{D,\ell^\pm} \equiv r^m_{D,\ell^+} + r^m_{D,\ell^-} .
\end{equation}
%


\subsection{Neutrino Fluxes}
\label{subsec:fluxes}

\begin{figure*}
    \includegraphics[width=0.48\textwidth]{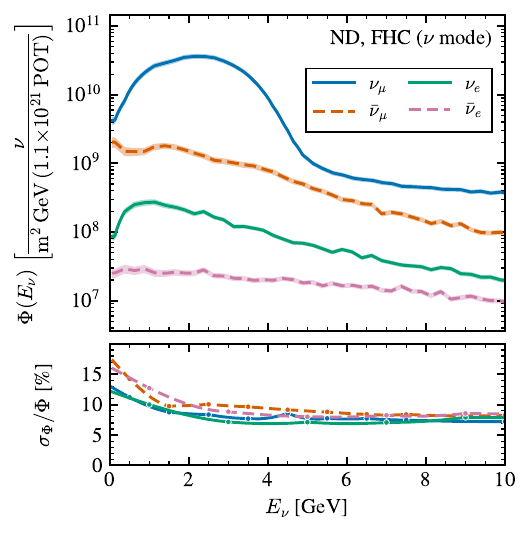}
    \hfill
    \includegraphics[width=0.48\textwidth]{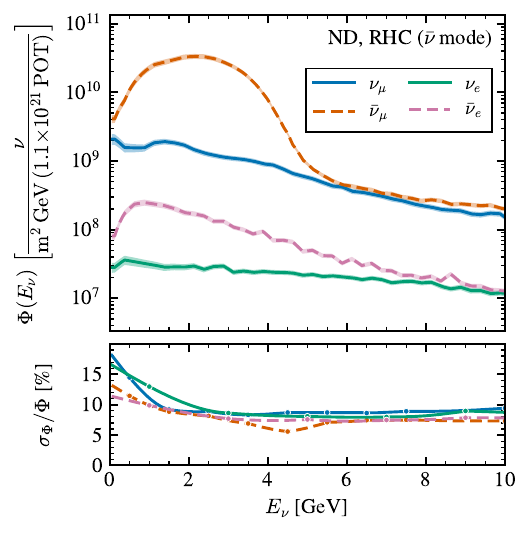}
    \caption{The DUNE ND neutrino fluxes from Refs.~\cite{DUNE:2020ypp,FieldsDune}, along with the fractional uncertainties used throughout this paper. The left side shows the FHC mode, while the right side shows the RHC mode.}
    \label{fig:ND_fluxes}
\end{figure*}

\begin{figure*}
    \includegraphics[width=0.48\textwidth]{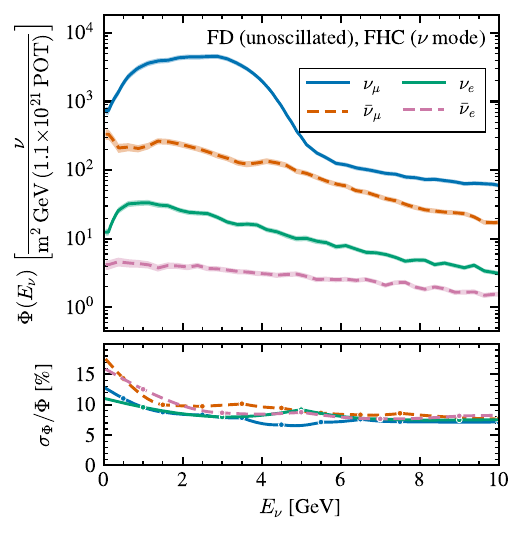}
    \hfill
    \includegraphics[width=0.48\textwidth]{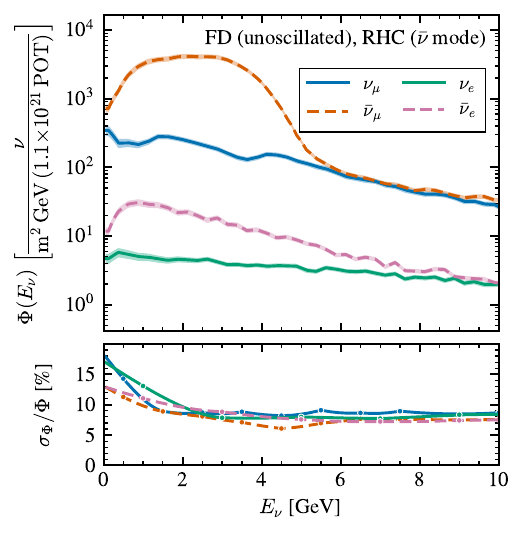}
    \caption{The unoscillated DUNE FD neutrino fluxes from Refs.~\cite{DUNE:2020ypp,FieldsDune}, along with the fractional uncertainties used throughout this paper. The left side shows the FHC mode, while the right side shows the RHC mode.}
    \label{fig:FD_fluxes}
\end{figure*}

In order to obtain the appropriate distributions of neutrino energies, we use the official DUNE fluxes for both the ND and FD~\cite{DUNE:2020ypp,FieldsDune}. The ND fluxes for FHC and RHC can be found in the left and right panels of Fig.~\ref{fig:ND_fluxes}, respectively.
The unoscillated FD fluxes for the FHC and RHC can be found in the left and right panels of Fig.~\ref{fig:FD_fluxes}, respectively.
As we can see, the FHC fluxes are dominated by the muon neutrino, with mild contamination from anti-muon neutrinos.
On the other hand, the RHC fluxes are dominated by the anti-muon neutrino, with significant contamination from the muon neutrinos.

In addition to the fluxes, we also include the uncertainties from the fluxes through the use of the covariance matrix from Refs.~\cite{DUNE:2020ypp,FieldsDune,IanKotler}. This matrix, reproduced in App.~\ref{app:flux_covariance},
includes correlations between the ND and FD fluxes, as well as between FHC / RHC modes and between lepton channels.
The ND and FD blocks of this full covariance matrix are strongly but not perfectly correlated: ND and FD bins with the same energy, mode, and channel have a median correlation of $0.94$ with a range that spans $0.68$--$0.99$. 
These correlations are included in all analyses below.

To generate flux realizations consistent with these uncertainties, we use the eigendecomposition of the full covariance matrix,
\begin{equation}
\label{eq:eigen_cov}
 \mathcal{C}^{mm'}_{DD'; \ell \ell'}(E_\nu, E_\nu') = \sum_{j=1}^{N_{\rm bins}} V^{m}_{j D \ell}(E_\nu)\, \lambda_j\, V^{m'}_{j D' \ell'}(E_\nu'),
\end{equation}
where $\lambda_j > 0$ by positivity. The principal components
\begin{equation}
   f^{m}_{j D \ell}(E_\nu) = V^{m}_{j D \ell}(E_\nu) \sqrt{\lambda_j}
\end{equation}
satisfy
\begin{equation}
 \mathcal{C}^{mm'}_{DD'; \ell \ell'}(E_\nu, E_\nu') = \sum_{j=1}^{N_{\rm bins}} f^{m}_{j D \ell}(E_\nu)\, f^{m'}_{j D' \ell'}(E_\nu').
\end{equation}
Covariance-matrix-weighted noise can thus be generated by rescaling unit normal random variables $a_j \sim \mathcal{N}(0,1)$ with $j \in \{1,\ldots,N_{\rm bins}\}$ by each principal component.
This allows noisy flux realizations
\begin{equation}
\label{eq:noisy_flux}
    \tilde{\Phi}_{D,\ell}^m(E_\nu; a_j) = \Phi_{D,\ell}^m(E_\nu) \left[ 1 + \sum_j a_j f^{m}_{j D \ell}(E_\nu) \right],
\end{equation}
to be realized with all cross-detector, cross-channel and cross-mode correlations included.
For ND training and FD oscillation analyses, we linearly interpolate the per-channel fluxes to obtain continuous functions of $E_\nu$ for each choice of $D,\,\ell,\,m$.


\subsection{Near detector training}
\label{sec:methods_ND}

Data-driven oscillation analyses rely on ND training to construct a cross-section model.
Here, we follow Ref.~\cite{Hackett:2024dul} and parameterize the SFs $W_i(x,Q^2)$ appearing in Eqs.~\eqref{eq:nuAd2} and~\eqref{eq:nuAd2anti} as fully connected two-dimensional NNs that can be meaningfully constrained by the two-dimensional flux-averaged ND rates $r^m_{D,\ell}(E_\ell,\cos\theta)$. Two important differences arise here in comparison with the single-channel muon disappearance studies of Ref.~\cite{Hackett:2024dul}. First, all five SFs $W_i(x,Q^2)$ must be treated as independent functions because $m^2_\ell/Q^2$ varies significantly between muons and electrons and therefore $W_4$ and $W_5$ enter differently from other SFs.
Second, for the case of a non-isoscalar target such as argon, $W^+$ and $W^-$ exchange couple to non-trivially different hadron tensors and the five antineutrino SFs $\overline{W}_i(x,Q^2)$ are parameterized by a separate NN.

The outputs of these two NNs, denoted $\widetilde{W}_i(x,Q^2)$ and $\widetilde{\overline{W}}_i(x,Q^2)$ with the NN parameters implicit, provide neutrino and antineutrino cross-section models
\begin{equation}
    \widetilde{\frac{d^2 \sigma^{(\nu A)}}{d\El d\costheta}}(E_\nu) = \sum_{i=1}^5 K_i(E_\nu,E_\ell,\cos\theta)\widetilde{W}_i(x,Q^2),
\end{equation}
\begin{equation}
    \widetilde{\frac{d^2 \sigma^{(\bar{\nu} A)}}{d\El d\costheta}}(E_{\bar{\nu}}) = \sum_{i=1}^5 \overline{K}_i(E_\nu,E_\ell,\cos\theta) \widetilde{\overline{W}}_i(x,Q^2).
\end{equation}
Note that $x$ and $Q^2$ are implicit functions of $E_\nu$ as well as $E_\ell$ and $\cos\theta$.
Neural models for the ND event rates are then obtained by marginalizing over the ND flux,
\begin{equation}
\label{eq:ND_model_-}
    q^m_{\text{ND},\ell^-}(E_\ell,\cos\theta) = \int dE_\nu \Phi^m_{\text{ND},\ell^-}(E_\nu) \widetilde{\frac{d^2 \sigma^{(\nu A)}}{d\El d\costheta}}(E_\nu),
\end{equation}
\begin{equation}
\label{eq:ND_model_+}
    q^m_{\text{ND},\ell^+}(E_\ell,\cos\theta) = \int dE_\nu \Phi^m_{\text{ND},\ell^+}(E_\nu) \widetilde{\frac{d^2 \sigma^{(\bar{\nu} A)}}{d\El d\costheta}}(E_\nu).
\end{equation}
The goal of ND training is to construct $q^m_{\text{ND},\ell^\pm}(E_\ell,\cos\theta)$ that accurately reproduces measured event-rate data, $r^m_{\text{ND},\ell^\pm}(E_\ell,\cos\theta)$.

As discussed above, ND training uses separate $\mu^-$ and $\mu^+$ rates but a combined $e^\pm$ rate. The NN prediction for the electron-like sample is
\begin{equation}
q^m_{\mathrm{ND},e^\pm}
= q^m_{\mathrm{ND},e^-} + q^m_{\mathrm{ND},e^+}.
\end{equation}
For numerical integration, we follow Ref.~\cite{Hackett:2024dul} and use the variables
\begin{equation}
\label{eq:v1v2}
    v_1 = \sqrt{E_\ell}, \qquad v_2 = \sqrt{E_\ell\sin^2(\theta/2)}.
\end{equation}
The kinematically allowed region is spread over a wider set of regular bins in these variables than in $(E_\ell,\cos\theta)$.
We construct the MSE out of the individual $\mu^\pm$ rates and the total $e^\pm$ rates,
\begin{equation}
\label{eq:MSE}
\begin{split} 
\mathrm{MSE} &= \sum_m \displaystyle\int dv_1 dv_2  \Bigl[ \left(r_{\text{ND},\mu^+}^m - q_{\text{ND},\mu^+}^m  \right)^2 \\
&\hspace{63pt} + \left(r_{\text{ND},\mu^-}^m - q_{\text{ND},\mu^-}^m  \right)^2 \\
&\hspace{63pt} + \left(r_{\text{ND},e^\pm}^{m} - q_{\text{ND},e^\pm}^{m}  \right)^2 \Bigr] ,
\end{split}
\end{equation}
where
\begin{equation}
    r^m_{D,\ell}(v_1,v_2) = J(v_1,v_2) \, r^m_{D,\ell}(E_\ell,\cos\theta),
\end{equation}
and $q^m_{D,\ell}(v_1,v_2)$ is defined analogously. The Jacobian is $J(v_1,v_2) = \left|\partial(E_\ell,\cos\theta)/\partial(v_1,v_2)\right| = 8v_2/v_1$.
Including both FHC and RHC modes, the MSE includes six contributions.
Exploration of more complicated loss functions involving different relative weights of the six terms is left to future work.

Training is accelerated by decomposing the SFs into pieces that separately govern their shapes and their normalizations, and then solving for the optimal normalization parameters using variable projection (VarPro) techniques.
Concretely, the structure functions are expressed as
\begin{equation}
    \widetilde{W}_i(x,Q^2) = c_i \, \text{NN}_i(x,Q^2),
\end{equation}
where $\text{NN}_i(x,Q^2)$ denotes the network output corresponding to the $i$-th SF and the $c_i$ are scalar prefactors. 
For neutrinos we set $i \in \{1,\ldots,5\}$ and for antineutrinos we take $i \in \{6,\ldots,10\}$. 
To define the VarPro scheme,
label each bin of the modeled ND event rate $q^m_{\text{ND},s}(v_1,v_2)$ by $a=(m,s,I)$, where $m$ is the beam mode, $s\in\{\mu^-,\mu^+,e^\pm\}$ is the observed lepton sample, and $I$ labels a bin in $(v_1,v_2)$. 
We denote the rate in that bin by $q_a$.
For fixed NN parameters, let $B_{ai}$ be the contribution of $\text{NN}_i$ to $q_a$ with unit coefficient $c_i$.
\begin{equation}
    B_{ai} = \int dE_\nu \, \Phi^m_{\text{ND},s}(E_\nu) \, K_i(E_\nu,I) \,\text{NN}_i(x,Q^2)
\end{equation}
The predicted rates and the loss can then be written as
\begin{equation}
    q_a = \sum_{i=1}^{10} B_{ai}c_i, \qquad
    \mathrm{MSE} = \sum_a (r_a-q_a)^2.
\end{equation}
The coefficients minimizing the MSE for fixed NN parameters are
\begin{equation}
    \mathbf{c}^{\star} = (B^T B)^{-1}B^T\mathbf{r},
\end{equation}
where $B$ has one row for each observed bin and one column for each of the ten SFs.

With this VarPro approach, nonlinear optimization is used only for the SF shape parameters, and the optimal normalization parameters for fixed shapes are computed within each step of the nonlinear optimization loop.
In particular, this avoids the need to manually restrict the NN output ranges to the same order of magnitude as the experimental rates without wasting many epochs correcting a poor initialization scale.

Details of our numerical implementation of this ND training scheme are discussed in Sec.~\ref{sec:ND_results}.


\subsection{Far detector oscillation analysis}
\label{sec:methods_FD}

At the FD, neutrino oscillations change the flavor composition of the beam. 
Each FD flavor therefore receives contributions from the initial flavors, weighted by their oscillation probabilities.
We use $\phi^{m}_{\text{FD}}$ to denote the unoscillated flux at the FD, so the oscillated FD flux can be written as
\begin{equation}
\label{eq:FD_flux}
    \Phi^m_{\text{FD},\ell}(E_\nu; \omega)
        = \sum_{\ell'} \phi^{m}_{\text{FD},\ell'}(E_\nu) \, 
        P_{\ell' \rightarrow \ell}(E_\nu; \omega),
\end{equation}
where $P_{\ell' \rightarrow \ell}$ is the probability for a neutrino produced as $\nu_{\ell'}$ to be in flavor state $\nu_\ell$ at the FD. 
The terms with $\ell=\ell'$ describe disappearance channels, and those with $\ell\neq\ell'$ describe appearance channels. 
The parameters $\omega$ include the mixing angles and CP-violating phase of the PMNS matrix and the mass splittings $\Delta m^2_{ij}$.

The inclusive FD charged lepton event rates are given by integrals of these fluxes and the relevant cross section, Eq.~\eqref{eq:rate}.
The NN cross-section model introduced above provides an approximation to this event rate,
\begin{equation}
\label{eq:FD_model_-}
    q^m_{\text{FD},\ell^-}(E_\ell,\cos\theta;\omega) = \int dE_\nu \Phi^m_{\text{FD},\ell^-}(E_\nu;\omega) \widetilde{\frac{d^2 \sigma^{(\nu A)}}{d\El d\costheta}}(E_\nu),
\end{equation}
\begin{equation}
\label{eq:FD_model_+}
    q^m_{\text{FD},\ell^+}(E_\ell,\cos\theta;\omega) = \int dE_\nu \Phi^m_{\text{FD},\ell^+}(E_\nu;\omega) \widetilde{\frac{d^2 \sigma^{(\bar{\nu} A)}}{d\El d\costheta}}(E_{\nu}).
\end{equation}
As discussed above, the FD analysis uses charge-inclusive muon-like and electron-like samples. Their predicted rates are
\begin{equation}
    q^m_{\text{FD},\ell^\pm} = q^m_{\text{FD},\ell^+} + q^m_{\text{FD},\ell^-},
\end{equation}
The corresponding total rate is
\begin{equation}
    N^m_{\ell,\mathrm{tot}}(\omega) = \int dE_\ell\,d\cos\theta\;q^m_{\text{FD},\ell^\pm}(E_\ell,\cos\theta;\omega).
\end{equation}
In the closure tests below, $\Phi^m_{\text{FD},\ell}$ is rescaled by the exposure required to achieve a fiducial number of events, and therefore $N^m_{\ell,\mathrm{tot}}(\omega)$ represents the total number of counts.

For each lepton flavor $\ell\in\{e,\mu\}$ and beam mode $m$, the FD data consist of $n_\ell^m$ events with measured kinematics $(E_\ell^{m(i)},\cos\theta^{m(i)})$. 
We fit their distributions, both shape and absolute rates, using the unbinned Poisson likelihood
\begin{equation}
  \begin{split}
    \mathcal{L}(\omega) &= \prod_{m,\ell} \text{Pois}(n_\ell^m | N^m_{\ell,\text{tot}}(\omega)) \\
    &\hspace{20pt}\times \prod_{i=1}^{n_\ell^m}
    \frac{q^m_{\text{FD},\ell^\pm}(E_\ell^{m(i)},\cos\theta^{m(i)};\omega)}
    {N^m_{\ell,\mathrm{tot}}(\omega)} \\
    &\propto \prod_{m,\ell} e^{-N^m_{\ell,\text{tot}}(\omega)} \prod_{i=1}^{n_\ell^m}
    q^m_{\text{FD},\ell^\pm}(E_\ell^{m(i)},\cos\theta^{m(i)};\omega),
  \end{split}
\end{equation}
where $\omega$-independent constants are omitted in the last line.
This differs from the shape-only likelihood of Ref.~\cite{Hackett:2024dul} by the Poisson fluctuations of the total rates in the first line.
The best-fit oscillation parameters are
\begin{equation}
    \omega^* = \mathrm{argmax}_\omega \mathcal{L}(\omega) ~ .
\end{equation}
Maximization of the likelihood $\mathcal{L}(\omega)$ is made numerically challenging by the presence of multiple local maxima, for instance solutions that differ in the octant of $\theta_{23}$.
Multi-start methods including seeds on opposite sides of known degeneracy points are used to mitigate these challenges below as detailed in Sec.~\ref{sec:FD_results}.

We emphasize that as in Ref.~\cite{Hackett:2024dul}, the NN parameters defining the neural cross-section model are held fixed during the FD analysis, and only the neutrino oscillation parameters are varied.
Inaccuracies in the trained neural model lead to systematic uncertainties in oscillation results explored for the muon disappearance channel in Ref.~\cite{Hackett:2024dul} and for a multi-channel appearance/disappearance analysis in the closure test below.


\section{Closure test}
\label{sec:closure_test}

For the closure test, we follow the analysis strategy of Ref.~\cite{Hackett:2024dul}, now including both electron and muon channels in FHC and RHC modes. This gives eight separate event samples, $e_m, \bar{e}_m, \mu_m, \bar{\mu}_m$, where $m$ represents either FHC or RHC mode. As discussed in Sec.~\ref{sec:methods_FD}, the oscillation analysis combines the two charges for each lepton flavor.

We obtain the FD fluxes by oscillating the unoscillated fluxes as described in Sec.~\ref{sec:methods_FD}. 
We take the true neutrino oscillation parameters from the NuFit-6.1 fit~\cite{Esteban:2024eli,NuFIT:6.1} using the normal ordering: $\sin^2\theta_{23}=0.470$, $\sin^2\theta_{12}=0.3088$, $\sin^2\theta_{13}=0.02249$, $\Delta m_{21}^2 = 7.537\times 10^{-5}$ eV${}^2$, $\Delta m_{31}^2 = 2.521\times 10^{-3}$~eV${}^2$, and $\delta_\text{CP} = 207^\circ$.
The oscillations are calculated, including matter effects, using the NuFast package~\cite{Denton:2024pzc} with the default matter density $\rho = 3 \text{ g/cm}^3$.
To set the FD sample sizes, we fix the $\mu^-$ event count in FHC mode to the DUNE TDR value of 6200 events in the region $[0.5, 8]$ GeV~\cite{DUNE:2020ypp}. This fixes the exposure for our toy model, taken equal for FHC and RHC running, and therefore the absolute scales of $\Phi^m_{\text{FD},\ell}(E_\nu)$. 
We then use the same exposure, together with the toy cross section and oscillated fluxes, to determine the event counts in all the other channels. The resulting sample sizes are shown in Table~\ref{tab:events}. Because our toy cross section differs from the models used for the DUNE projections, the resulting event counts in the other channels need not match the corresponding DUNE predictions.

\begin{table}[t]
\centering
\renewcommand{\arraystretch}{1.25}
\begin{tabular}{@{\hspace{6pt}}l@{\hspace{12pt}}|@{\hspace{12pt}}c@{\hspace{12pt}}|@{\hspace{12pt}}c@{\hspace{6pt}}}
\hline
FD sample & FHC & RHC \\
\hline
$\mu^-$ & 6200 & 1525 \\
$e^-$   & 1512 & 203  \\
$\mu^+$ & 955 & 4091 \\
$e^+$   & 103  & 493  \\
\hline
$\mu^\pm$ total & 7155 & 5616 \\
$e^\pm$ total   & 1615 & 696  \\
\hline
\end{tabular}
\caption{Numbers of simulated FD events in FHC and RHC modes. The FHC $\mu^-$ count is fixed to the DUNE TDR value; all other counts follow from the same exposure with the toy cross section and oscillated fluxes. Charge-inclusive totals are used in the oscillation analysis.}
\label{tab:events}
\end{table}


\subsection{Cross-section toy model}

The true cross section model used in this closure test is a minor extension of the one used in Ref.~\cite{Hackett:2024dul}.
The SFs are calculated using the leading-order deep inelastic scattering results from the quark-parton model~\cite{Bjorken:1969ja}. 
For the neutrino-nucleus charged current reaction, the structure functions are 
\begin{align}
    W_2 =&  \frac{4x^2 M_A^2}{A Q^2}\left(Z\left( f_{\bar{u}/p} + f_{d/p}+f_{\bar{c}/p}+f_{s/p}\right) \right. \nonumber \label{eq:W2_neutrino} \\
    &\left. +(A-Z)\left(f_{\bar{u}/n} + f_{d/n}+f_{\bar{c}/n}+f_{s/n}\right)\right)\,, \\
    W_3 =&  2x\left(Z\left( f_{d/p} -f_{\bar{u}/p} +f_{s/p}- f_{\bar{c}/p}\right)\right. \nonumber \label{eq:W3_neutrino} \\
    &\left. +(A-Z)\left( f_{d/n} -f_{\bar{u}/n} +f_{s/n}- f_{\bar{c}/n}\right)\right)\,,
\end{align}
where $f_{i/N}$ is the parton distribution function (PDF) for a parton $i$ in a nucleon $N$, $Z$ is the number of protons in the nucleus, and $A$ is the total number of nucleons. The dependence of $x$ and $Q^2$ in the PDF is suppressed for brevity. 
The neutron PDFs are obtained from isospin symmetry, and in practice involve the relationship $f_{u/p} = f_{d/n}$, $f_{\bar{u}/p} = f_{\bar{d}/n}$, and similar for the down quarks in the proton. The strange- and charm-quark PDFs and the gluon PDF are taken to be the same in the proton and neutron.
For the antineutrino-nucleus charged current reaction, the SFs are 
\begin{align}
    \overline{W}_2 =& \frac{4x^2 M_A^2}{A Q^2}\left(Z\left( f_{\bar{d}/p} + f_{u/p}+f_{\bar{s}/p}+f_{c/p}\right) \right. \nonumber \label{eq:W2_aneutrino} \\
    &\left. +(A-Z)\left(f_{\bar{d}/n} + f_{u/n}+f_{\bar{s}/n}+f_{c/n}\right)\right)\,, \\
    \overline{W}_3 =& 2x\left(Z\left( f_{u/p} -f_{\bar{d}/p} +f_{c/p}- f_{\bar{s}/p}\right)\right. \nonumber \label{eq:W3_aneutrino} \\
    &\left. +(A-Z)\left( f_{u/n} -f_{\bar{d}/n} +f_{c/n}- f_{\bar{s}/n}\right)\right)\,.
\end{align}
For both neutrinos and antineutrinos, $W_1$ is obtained from the Callan--Gross relation ($2xW_1 = \frac{AQ^2}{2M_A^2}W_2$)~\cite{Callan:1969uq}, and $W_4$ and $W_5$ are given by the tree-level relations ($2xW_5=\frac{AQ^2}{2M_A^2}W_2$, $W_4=0$)~\cite{Albright:1974ts}.

For this study, we use the CT18NNLO PDF set~\cite{Hou:2019efy}. The PDFs are evaluated through the use of the python interface in LHAPDF~\cite{Buckley:2014ana}. 
Since the region of $x$ and $Q^2$ relevant for DUNE probes regions in which the PDF is not well-defined, we extrapolate smoothly into this region using the MSTW prescription~\cite{Martin:2009iq}.
To ensure that the extrapolation does not violate cross-section positivity for either neutrinos or antineutrinos, we rescale all SFs by $Q^2/(Q^2 + 1\text{ GeV}^2)$ and further rescale $W_3$ / $\overline{W}_3$ by a factor of 1/10.
The precise details of the toy model are not expected to have a significant impact on the physics conclusions of this work.

Using Eqs.~\eqref{eq:W2_neutrino} and~\eqref{eq:W3_neutrino}, and the relations for $W_1, W_4,$ and $W_5$ in Eq.~\eqref{eq:nuAd2} defines the true cross section model for neutrino-nucleus interactions in the closure test.
Similarly, Eqs.~\eqref{eq:W2_aneutrino} and~\eqref{eq:W3_aneutrino} are used in conjunction with Eq.~\eqref{eq:nuAd2anti} to obtain the true cross section model for antineutrino-nucleus interactions in the closure test.
Finally, the closure test uses argon-40 ($Z=18$, $A=40$) as the nucleus for the calculation, since the main focus is on the DUNE experiment.


\begin{figure}
    \centering
    \includegraphics[width=\linewidth]{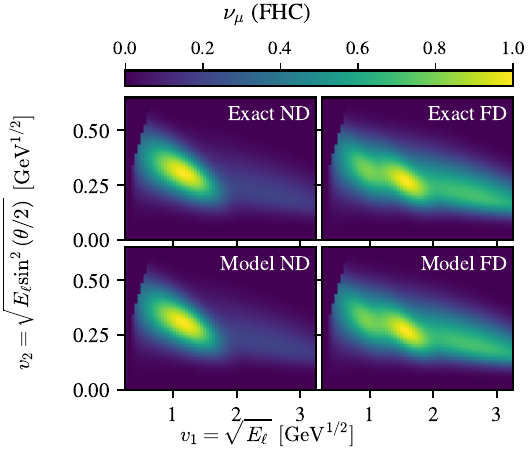}\vspace{10pt}
    \includegraphics[width=\linewidth]{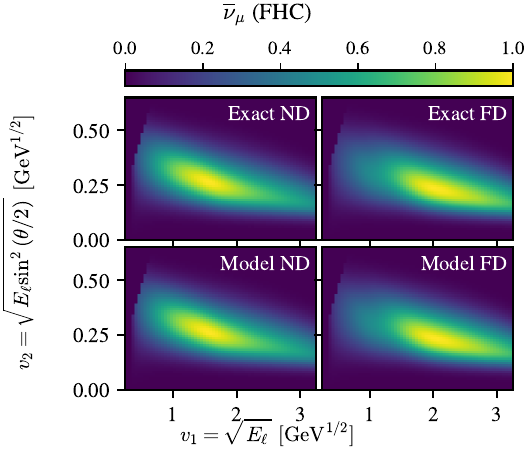}\vspace{10pt}
    \includegraphics[width=\linewidth]{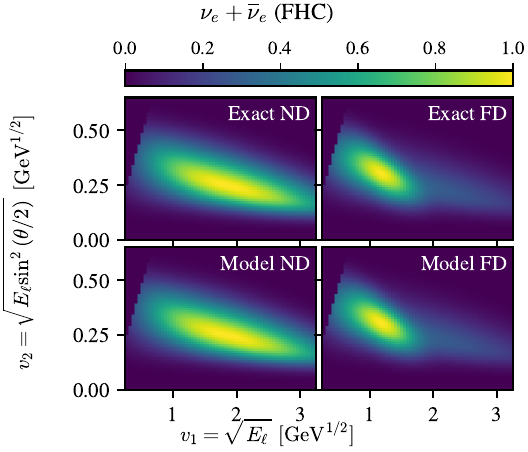}
    \caption{Near-detector (left) and far-detector (right) event rates in FHC mode predicted using the true (top) or learned (bottom) cross sections and the true ND and FD fluxes. Event rates for $\mu^+$ and $\mu^-$ are considered separately while $e^\pm$ rates are added jointly.}
    \label{fig:ND_FHC}
\end{figure}

\begin{figure}
    \centering
    \includegraphics[width=\linewidth]{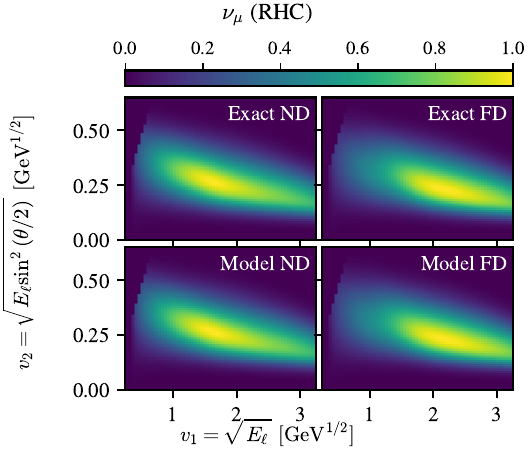}\vspace{10pt}
    \includegraphics[width=\linewidth]{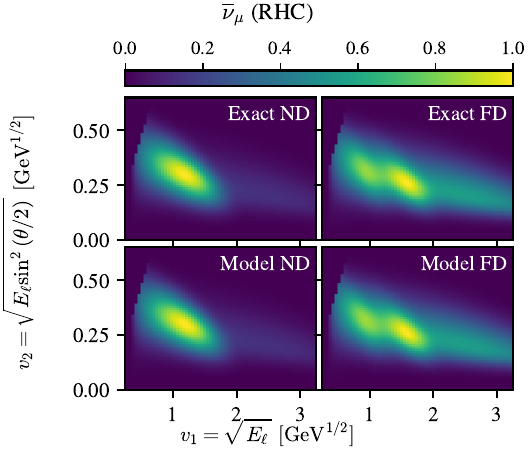}\vspace{10pt}
    \includegraphics[width=\linewidth]{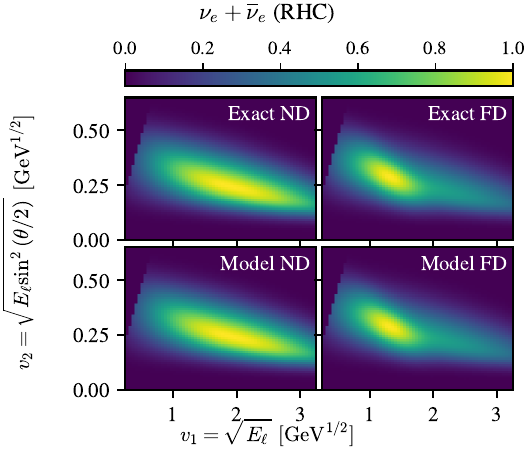}
    \caption{Near-detector (left) and far-detector (right) event rates in RHC mode predicted using the true (top) or learned (bottom) cross sections and the true ND and FD fluxes. Event rates for $\mu^+$ and $\mu^-$ are considered separately while $e^\pm$ rates are added jointly.
    \label{fig:ND_RHC}}
\end{figure}

\subsection{Near detector results}
\label{sec:ND_results}

Simulated ND training data in the form of event-rate $r^m_{\text{ND},\ell}(E_\ell,\cos\theta)$ histograms are obtained from the toy model above by directly evaluating Eq.~\eqref{eq:rate} for each histogram bin using DUNE flux estimates $\Phi_{\text{ND},\ell}^m$ described in Sec.~\ref{subsec:fluxes} for $m \in \{\text{FHC},\text{RHC}\}$ and $\ell \in \{\mu^+,\mu^-,e^+,e^-\}$.
We first focus on the case of infinite ND statistics with a perfectly known flux and no detector smearing to establish an idealized baseline where finite FD statistics are the only limitation.
Effects of finite ND statistics, flux and other experimental systematics, and NN initialization systematics are deferred to Sec.~\ref{sec:systematics} below.

\begin{figure*}
    \centering
    \includegraphics[width=\linewidth]{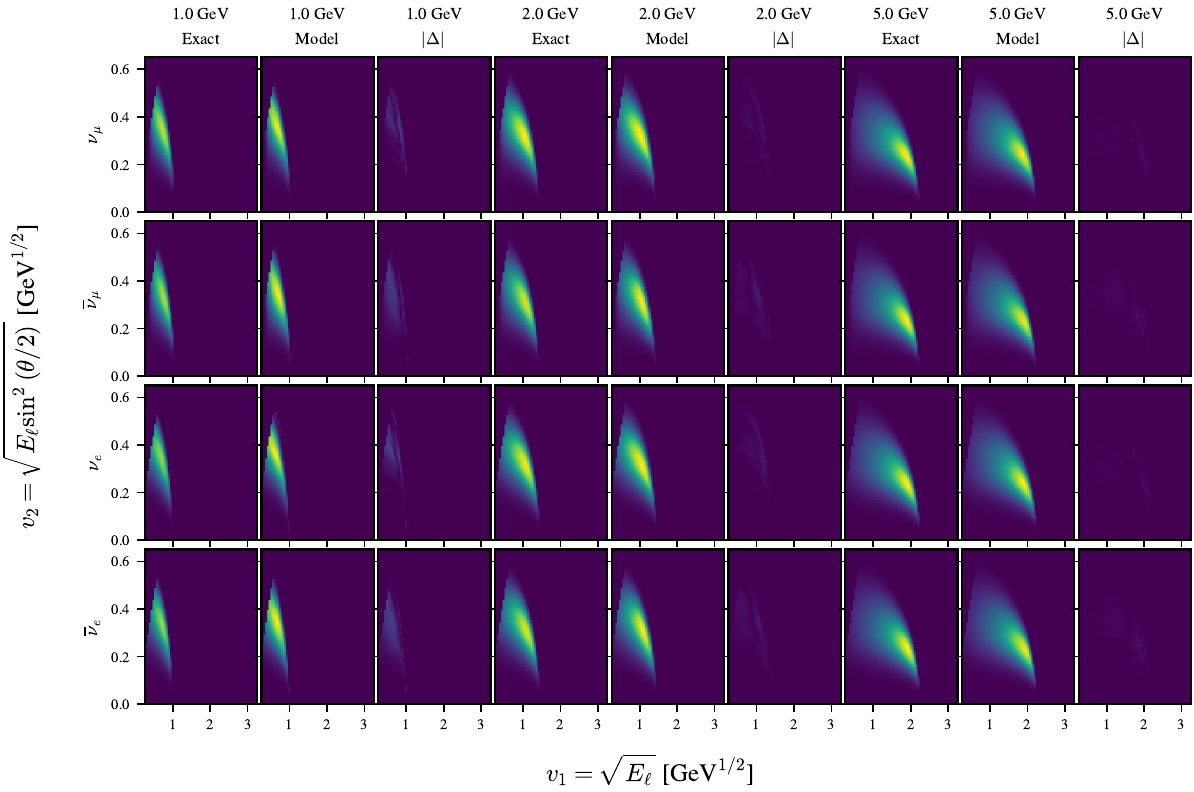}
    \caption{ Comparisons of true and learned cross sections for each lepton flavor on three slices of fixed $E_{\nu}$. Note that results for each lepton and $E_{\nu}$ slice are normalized separately to highlight the level of shape agreement; see Fig.~\ref{fig:total_1d} for overall scale comparisons. }
    \label{fig:marginal_3d}
\end{figure*}

Concretely, we use two fully connected NNs (or multi-layer perceptrons), one for neutrino SFs and one for antineutrino SFs. Each has two inputs, $x$ and $Q^2$, five SF outputs, corresponding to $W_i(x,Q^2)$ and $\overline{W}_i(x,Q^2)$, four hidden layers of width 64, and leaky ReLU activation functions.
Training used a $16\times 16$ grid of $v_1/\sqrt{\text{GeV}} \in [0.25, 3.25]$  and $v_2/\sqrt{\text{GeV}} \in [0, 0.65]$ for the $(v_1,v_2)$ variables defined in Eq.~\eqref{eq:v1v2} with $E_\nu$ integrals performed on a 32-point grid with $E_\nu \in [m_{\mu}, 10\text{ GeV}]$.
We trained for $10^5$ epochs using the Adam optimizer~\cite{Kingma:2014vow}, gradient clipping with norm unity, and VarPro methods to determine the optimal amplitudes $c_i$ for each structure function as described in Sec.~\ref{sec:methods_ND}.
The training loss is taken to be the MSE in Eq.~\eqref{eq:MSE} times a constant chosen to make the pre-training loss $O(1)$; without this rescaling neutrino event rates in conventional units are smaller than default Adam numerical regularization parameters and training is pathological.
We used an initial learn rate (LR) of $10^{-3}$ and a plateau LR scheduler with a reduction factor of 0.5, improvement threshold of $10^{-2}$, patience of 2000, and LR floor of $10^{-5}$.
Typically, the LR floor is reached after 60,000--75,000 training steps.
In comparison with the training strategy used in Ref.~\cite{Hackett:2024dul}, we use ten times more training epochs but eight-fold coarser resolution in each $(v_1,v_2,E_\nu)$ dimension, which we found to be more computationally efficient for achieving a given level of training quality in terms of total compute cost.
Training each network initialization took approximately 15--20 minutes on a single A100 GPU.

The ND event rates for each channel used as training data are compared to the trained NN predictions in Figs.~\ref{fig:ND_FHC} and~\ref{fig:ND_RHC}.
The agreement is good, as expected for a network trained to minimize exactly this difference.
The critical comparison is the FD event rates obtained by multiplying the toy/learned model cross sections by the (NuFit 6.1 ``truth'') oscillation probabilities before flux averaging, which is also shown in Figs.~\ref{fig:ND_FHC} and~\ref{fig:ND_RHC}.
Clear agreement is seen in this case, which shows that the learned model extrapolates correctly to a different $E_\nu$ flux.
As demonstrated in Ref.~\cite{Hackett:2024dul}, extrapolation between ND and FD fluxes does not work for black-box three-dimensional $\frac{d^2\sigma}{dE_\ell d\cos\theta}$ models. 
The key feature of SF parameterizations of the cross section is that an accurate generalization from ND to FD data only requires the two event distributions to overlap in $(x,Q^2)$.

Detailed comparisons of true/learned cross sections are shown for each one-dimensional marginal of the three-dimensional distributions $\Phi_{\text{ND},\ell}^m \frac{d^2\sigma^{(\nu A)}}{dE_\ell d\cos\theta}(E_\nu)$ in Fig.~\ref{fig:marginal_1D_ND}.
Analogous comparisons for the FD distributions $\Phi_{\text{FD},\ell}^m \frac{d^2\sigma^{(\nu A)}}{dE_\ell d\cos\theta}(E_\nu)$ are shown in Fig.~\ref{fig:marginal_1D_FD} and demonstrate that the learned cross section is accurately modeling the FD distribution over its full kinematic range.
Further comparisons of true/learned double-differential cross sections at several fixed values of $E_\nu$ are shown in Fig.~\ref{fig:marginal_3d} and indicate that the cross-section shape is modeled accurately at fixed $E_{\nu}$, although some slight distortions are visible at lower $E_{\nu}$, in particular below the ND flux peak.
The energy dependence of the total cross section $\sigma(E_\nu)$ obtained by integration of the true/learned differential cross-section models is shown in Fig.~\ref{fig:total_1d} and indicates that the normalization as well as shape of the cross section is accurately learned for each lepton species.

It is noteworthy that accurate extrapolations to FD kinematics are obtained even for $e^\pm$ channels where we assume no information about charge-identification is available and train on the total $e^\pm$ event rate.
The presence of separate $\mu^+$ and $\mu^-$ event rates is essential for breaking degeneracies between the $W_i$ and $\overline{W}_i$ for $i \in \{1,2,3\}$, and with only total $e^\pm$ rate information available there can be expected to be degeneracies that are not broken between $m_\ell^2/Q^2$-suppressed structure functions.
The flux-averaged ND rates therefore do not determine every structure function individually; what makes the FD extrapolation work is the overlap of the ND and FD event distributions in $(x,Q^2)$.


\subsection{Far detector results}
\label{sec:FD_results}

We fit the simulated FD samples using either the true or learned cross section and the likelihood defined in Sec.~\ref{sec:methods_FD}. The event counts for each sample are given in Table~\ref{tab:events}.
With the true cross section, the maximum likelihood estimate (MLE) of the oscillation parameters $\omega_\star$ is expected to approach the input (NuFit 6.1) parameters $\omega_{\rm true}$ as FD statistics increase.

For one sample of simulated FD events, finite-statistics bias is present.
The difference between $\omega_{\rm true}$ and the $\omega_{\star}$ obtained using the true toy-model cross section provides a rough indication of the size of FD statistics effects in this toy model, but we emphasize that the constraining power of a realistic inclusive-only DUNE analysis could differ significantly from this.
The purpose of the closure test is to measure the additional bias and variance introduced by the neural model, by comparing the $\omega_\star$ obtained with the trained model to the $\omega_\star$ obtained with the true cross section on the same FD sample.
We have repeated the analysis below using different random seeds, different values of $\omega_{\rm true}$, and different FD analysis strategies.
These results demonstrate that differences between $\omega_{\star}$ and $\omega_{\rm true}$ fluctuate around zero as expected for finite-statistics effects and that fluctuations in the learned cross-section $\omega_{\star}$ closely track those of the true cross-section.

To compute confidence intervals (CIs) quantifying best-fit oscillation parameter regions,
the simulated FD events are bootstrap resampled~\cite{Efron:1982,DiCiccio:1996,Davison:1997} to form $N_{\rm boot} = 4096$ bootstrap ensembles.
Each ensemble's size is drawn randomly from a $\mathrm{Pois}(n^m_\ell)$ distribution so that the absolute rates carry their proper statistical fluctuations.
For each bootstrap sample $b$, the MLE oscillation parameters $\omega_{\star}^{(b)}$ are computed independently using an L-BFGS optimizer implemented in \texttt{PyTorch}~\cite{NEURIPS2019_9015}.
To avoid getting trapped in local minima, particularly those arising from the $\theta_{23}$ octant near-degeneracy, we use the global maximum of optimizer results obtained using 8 different initial seeds for each bootstrap sample: four random oscillation parameters in one $\theta_{23}$ octant plus their four octant mirrors.
Empirical bootstrap confidence intervals of the $\omega_{\star}^{(b)}$ are then used to define CIs.

\begin{figure}
    \centering
    \includegraphics[width=.95\linewidth]{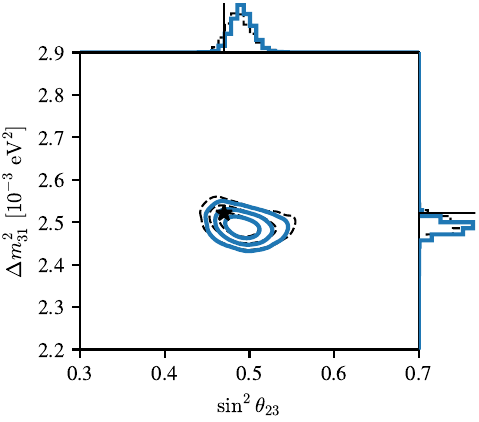}
    \includegraphics[width=.95\linewidth]{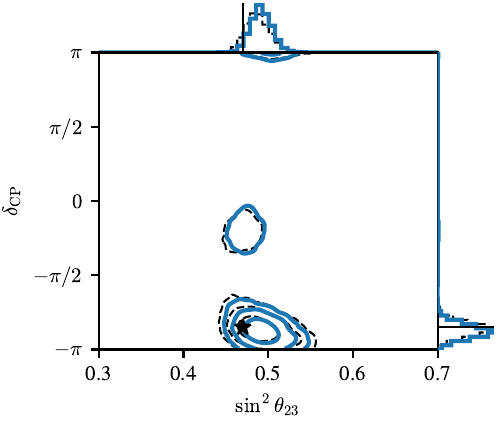}
    \includegraphics[width=.95\linewidth]{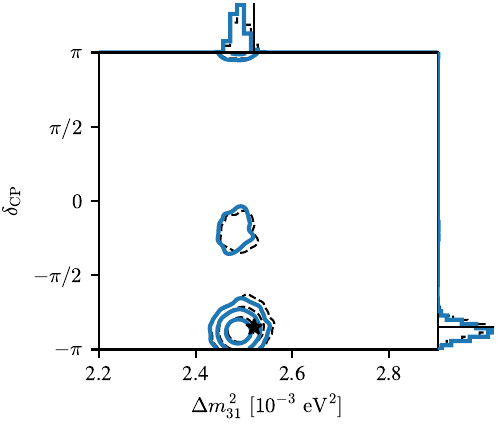}
    \caption{Confidence intervals inferred using the true cross section (dashed black) and learned model (solid blue). Contour lines are obtained from a kernel density estimate (KDE) constructed from the maximum-likelihood parameters of 4,096 bootstrap samples of FD events. Histograms show marginal distributions for each parameter. Stars and vertical lines indicate the true values used in this closure test. } 
    \label{fig:confused_nosys}
\end{figure}

The resulting confidence regions for the true and learned cross-section models are shown in Fig.~\ref{fig:confused_nosys}. The learned-model contours closely track those obtained with the true cross section for all three oscillation parameters. Both fits show the same octant ambiguity in $\theta_{23}$ and multiple allowed regions for $\delta_{\rm CP}$, with no apparent additional bias from the learned cross section. To verify that the learned $\omega_{\star}$ genuinely tracks the true cross section $\omega_{\star}$ as well as approximating $\omega_{\rm true}$, we repeated the entire analysis with a second FD statistics draw where finite-statistics bias is relatively large and found that the learned $\omega_{\star}$ indeed tracks the finite-statistics bias of the true cross-section $\omega_{\star}$.

\section{Experimental and training systematics}
\label{sec:systematics}

The closure test above demonstrates that a learned cross-section model can be used to accurately perform multi-channel oscillation analyses in an idealized setup in which the only uncertainties arise from finite FD statistics.
This section extends the closure test above to model systematic uncertainties associated with finite ND statistics, ND and FD detector smearing, flux uncertainties, and NN initialization.
Treatment of these systematic uncertainties largely parallels the single-channel treatment in Ref.~\cite{Hackett:2024dul}. However, flux uncertainties have more significant effects on multi-channel oscillation analyses than in the single-channel case.
For this reason, we adopt a more detailed treatment of flux systematics where flux nuisance parameters are included in ND fits as discussed below.


\subsection{Finite ND statistics}

The DUNE ND will have extremely large event samples, but not infinite.
To account for this, we compute the ND event sample sizes similarly to our treatment of the FD.
We fix the dominant channel, $\nu_\mu$ in FHC mode, to the DUNE ND estimate of $3\times10^{7}$ accepted $\nu_\mu$ CC events per year~\cite{DUNE:2021tad}; this gives us the effective ND exposure.
Then, with the same exposure, taken equal for FHC and RHC running, we compute the sample sizes of all other channels with our toy cross-section model.
The resulting ND sample sizes are listed in Table~\ref{tab:events_ND}.
The other channels are broadly comparable with the DUNE ND estimates~\cite{DUNE:2021tad}, although a precise match is not expected for our toy cross section.
Two-dimensional event-rate histograms $r_{\rm ND,\ell}^m$ for all channels are then formed for the same $(v_1, v_2)$ grids as the infinite-statistics case.
Training then proceeds exactly as described above.

\begin{table}[t]
\centering
\renewcommand{\arraystretch}{1.25}
\begin{tabular}{@{\hspace{6pt}}l@{\hspace{12pt}}|@{\hspace{12pt}}c@{\hspace{12pt}}|@{\hspace{12pt}}c@{\hspace{6pt}}}
\hline
ND sample & FHC & RHC \\
\hline
$\mu^-$ & $3.0\times10^{7}$ & $6.1\times10^{6}$ \\
$e^-$   & $6.7\times10^{5}$ & $3.5\times10^{5}$ \\
$\mu^+$ & $3.4\times10^{6}$ & $1.9\times10^{7}$ \\
$e^+$   & $2.2\times10^{5}$ & $3.9\times10^{5}$ \\
\hline
$\mu^\pm$ total & $3.3\times10^{7}$ & $2.5\times10^{7}$ \\
$e^\pm$ total   & $8.9\times10^{5}$ & $7.4\times10^{5}$  \\
\hline
\end{tabular}
\caption{Numbers of simulated ND events in FHC and RHC modes. Charge-inclusive totals are also shown.}
\label{tab:events_ND}
\end{table}

Results for oscillation parameter CIs obtained using six different random draws of ND events to construct $r_{\rm ND,\ell}^m$ are compared with infinite-statistics results in Fig.~\ref{fig:confused_stats}.
The finite-statistics results are nearly indistinguishable from the infinite-statistics ones, with only isolated differences in $3\sigma$ coverage intervals visible.
This suggests that the infinite-statistics approximation should be reliable for inclusive cross-section learning at DUNE.
Finite ND statistics effects are included in the total uncertainty estimates below by including random ND event draws leading to different realizations of $r_{\rm ND,\ell}^m$ in total systematics samples.


\subsection{Detector smearing}

At the ND, both the simulated data and the model predictions are represented as event-rate histograms. We apply the $4\%$ charged-lepton energy smearing described in Sec.~\ref{subsec:detector} to both histograms, at fixed scattering angle.

At the FD, the simulated data consist of individual events. We apply a Gaussian random shift to each event's charged-lepton energy, with a width of $4\%$ of its true energy. We also convolve the predicted event rate with the same energy-resolution model when evaluating the oscillation likelihood.

The effects of both ND and FD smearing on oscillation parameter CIs are shown in Fig.~\ref{fig:confused_smear}.
For DUNE's energy resolution, the effects of detector smearing lead to shifts in the CIs that are much smaller than their $1\sigma$ widths.
Similar sizes of variation are seen for learned cross-section analysis and for oscillation analyses using smeared versions of the true toy-model cross section, suggesting that detector-smearing effects on ND training are subdominant to smearing effects in FD likelihood calculations.
Both ND and FD detector-resolution systematics are included in the total uncertainty estimates below using this combination of random shifts to FD data samples and explicit convolutions of ND event rates and FD likelihoods with Gaussian smearing kernels.


\subsection{Flux uncertainties}

\begin{figure*}
    \centering
    \includegraphics[width=\linewidth]{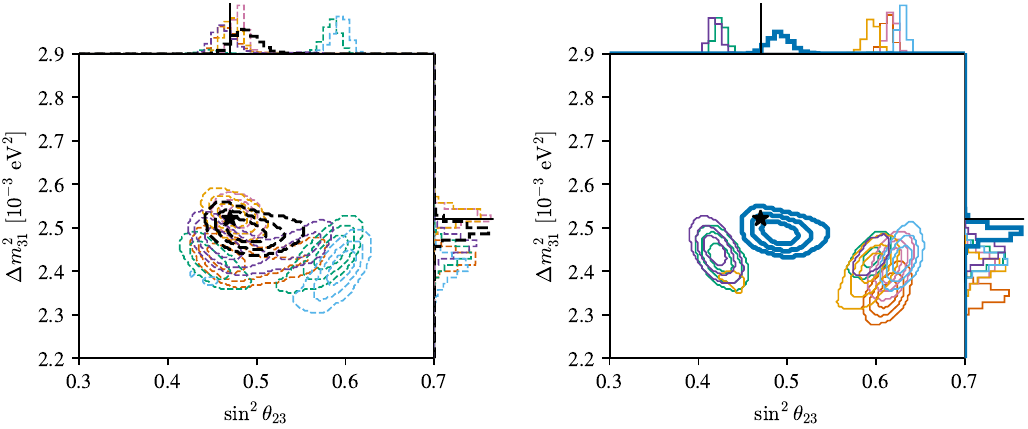}
    \includegraphics[width=\linewidth]{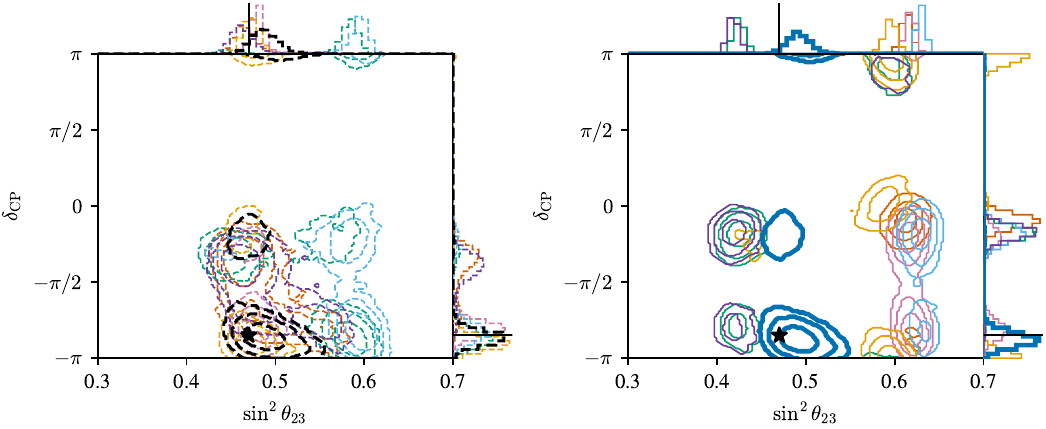}
    \includegraphics[width=\linewidth]{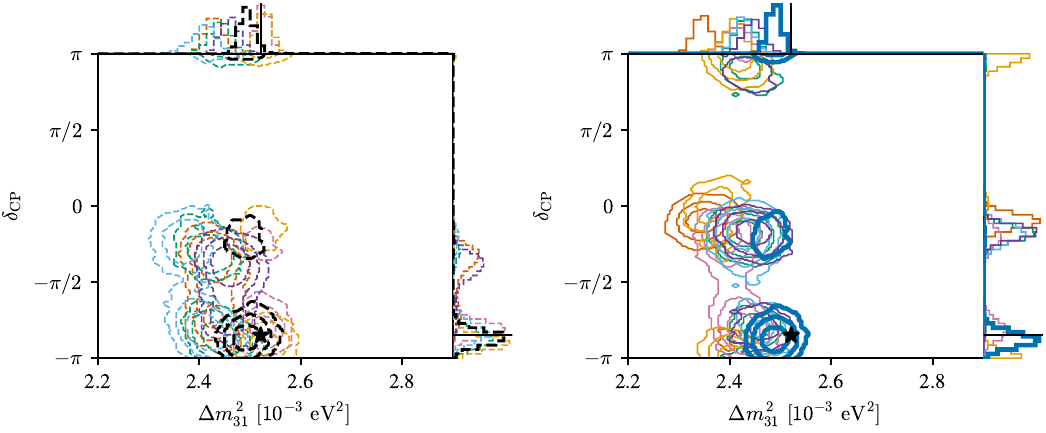}
    \caption{Confidence intervals for six different flux-model draws analogous to Fig.~\ref{fig:confused_smear} without including the forward-folded nuisance parameter fitting strategy described in the main text, for the true cross section (left, dashed colored lines) and the learned cross section (right, solid colored lines), compared with the corresponding results without flux uncertainties (dashed black and thick blue lines). Relatively large distortions arise from the mismatch between data and model $E_\nu$-dependence induced by flux uncertainties.}
    \label{fig:too_confused_flux}
\end{figure*}

\begin{figure*}
    \centering
    \includegraphics[width=\linewidth]{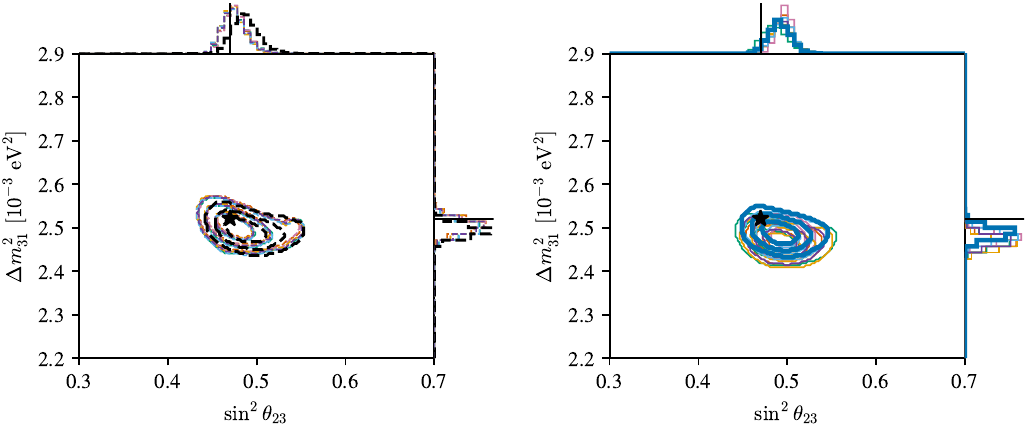}
    \includegraphics[width=\linewidth]{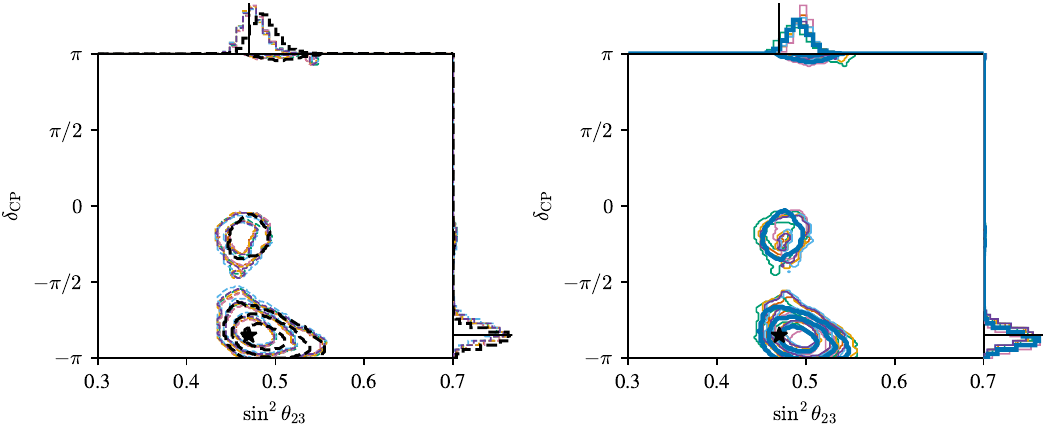}
    \includegraphics[width=\linewidth]{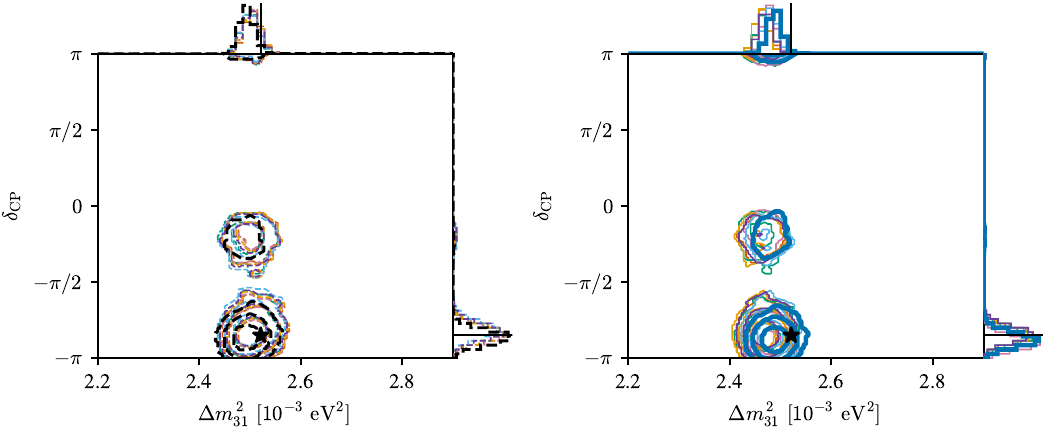}
    \caption{Confidence intervals for the same six flux-model draws as Fig.~\ref{fig:too_confused_flux} with the forward-folded nuisance parameter fitting strategy described in the main text.}
    \label{fig:confused_flux}
\end{figure*}

The total flux covariance matrix described in Sec.~\ref{subsec:fluxes} includes a total of $N_{\rm bins} = 136$ energy bins: 68 for the ND and 68 for the FD, each covering the eight beam modes and flavor channels.
A single consistent realization of ND and FD fluxes including correlated uncertainties is obtained by drawing a fixed set of unit normal variables $a_j^\star$ and applying it to both detector fluxes as in Eq.~\eqref{eq:noisy_flux},
\begin{equation}
  \tilde{\phi}_{D,\ell}^m(E_\nu;a_j^\star) = \phi_{D,\ell}^m(E_\nu) \left[ 1 + \sum_{j=1}^{N_\text{bins}} a_j^\star f^{m}_{j D \ell}(E_\nu) \right].
\end{equation}
To implement this procedure, we generate simulated ND data histograms from $\tilde{\phi}_{\text{ND},\ell}^m(E_\nu;a_j^\star)$, and a new set of FD events by using $\tilde{\phi}_{\text{FD},\ell}^m(E_\nu;a_j^\star)$ in place of the unoscillated flux $\phi_{\text{FD},\ell}^m(E_\nu)$ in Eq.~\eqref{eq:FD_flux}, again fixing $N_{\text{FD},\mu^-}^{\text{FHC}} = 6200$ with event rates for the other channels determined self-consistently.

Figure~\ref{fig:too_confused_flux} shows the results of the oscillation analysis with flux uncertainties included.
The systematic variations are large.
In particular, the $\Delta m^2_{31}$ CIs vary between flux draws by more than the FD statistical uncertainty and by more than any of the other experimental systematics discussed above.
The variations are also larger for the learned cross section than for the true one.
They are also much larger than the flux systematics found in the single-channel analysis of Ref.~\cite{Hackett:2024dul}.
Adding channels should not make a systematic effect worse, which suggests a problematic feedback between the flux uncertainty modeling and the ND training process.

The root cause of the problem is a mismatch between the energy-dependence of the actual data distribution, which involves $\widetilde{\Phi}(E_\nu)$, and that of the model distribution, which involves $\Phi(E_\nu)$.
Because the explicit $E_\nu$ dependence of the cross section is fixed by the kinematic coefficients, the network can only compensate for distortions in $E_\nu$ by modifying the normalizations and the $(x,Q^2)$ shapes of structure functions.
The network therefore absorbs the flux mismatch into the structure functions: their normalizations shift to match the ND rates in each channel, and their $(x,Q^2)$ shapes develop unphysical channel-dependent distortions. When these distorted structure functions are used at the FD, the distortions turn into biases in the oscillation parameters.

To isolate and resolve the problem, we adopt a ``forward-folded'' flux uncertainty modeling strategy in which the learned cross section model is augmented with flux nuisance parameters $b_j$ as
\begin{equation}
\begin{split}
  \widetilde{\frac{d^2\sigma}{dE_\ell d\cos\theta}} &= \left[ 1 + \sum_{j=1}^{N_\text{ND\,bins}} b_j g^{m}_{j \ell}(E_\nu) \right]\sum_{i=1}^5 K_i(E_\nu,E_\ell,\cos\theta) \\
    &\hspace{10pt} \times c_i \, NN_i(x,Q^2),
    \end{split}
\end{equation}
where the $g^m_{j \ell}(E_\nu)$ are obtained from the Cholesky decomposition of the rank $N_\text{ND\,bins}=68$ ND covariance matrix
\begin{equation}
 \mathcal{C}^{mm'}_{\text{ND}\, \text{ND}; \ell \ell'}(E_\nu, E_\nu') = \sum_{j=1}^{N_\text{ND\,bins}} g^{m}_{j \ell}(E_\nu)\, g^{m'}_{j \ell'}(E_\nu'),
\end{equation}
and the $NN_i(x,Q^2)$ form the same neural model used above with coefficients $c_i$ obtained with VarPro.
The $b_j$ are fitted together with the NN parameters during training. 
The nuisance parameters give the model enough freedom to absorb a flux shift of the size allowed by the covariance matrix, without any knowledge of the particular draw $a_j^\star$ used to generate the data. 
The inclusion of the nuisance parameters $b_j$ should not be interpreted as a simultaneous construction of a cross-section model and a flux model, but rather as a way of incorporating the covariance-matrix results of a well-determined flux model into the cross-section modeling process.
The sensitivity of the cross-section learning process to flux uncertainties shows that further research on precisely combining flux and cross-section uncertainty modeling will be important in the context of data-driven models.

The flux nuisance parameters are fitted to ND data only, but the ND and FD fluxes are correlated, meaning the base FD flux entering the fits must be determined from the appropriate conditional distribution.
After training, the model ND flux deviates from the nominal flux by
\begin{equation}
\label{eq:deltaND}
    \delta_{\text{ND},\ell}^m(E_\nu) = \frac{\tilde{\Phi}^m_{\text{ND},\ell}(E_\nu)}{\Phi^m_{\text{ND},\ell}(E_\nu)}\left[ 1 + \sum_{j=1}^{N_\text{ND\,bins}} b_j^\star\, g^{m}_{j \ell}(E_\nu) \right] - 1,
\end{equation}
where $\tilde{\Phi}_{\text{ND}}$ is the ND flux assumed by the model (see below) and $b_j^\star$ are the best-fit nuisance parameters.
Because the ND and FD fluxes are correlated through the covariance matrix of Eq.~\eqref{eq:eigen_cov}, a measured ND deviation implies an expected FD deviation.
Writing $\mathcal{C}_{\text{ND},\text{ND}}$, $\mathcal{C}_{\text{FD},\text{ND}}$, and $\mathcal{C}_{\text{FD},\text{FD}}$ for the detector blocks of this matrix, the FD deviation is modeled as
\begin{equation}
\label{eq:deltaFD}
    \delta_{\text{FD}} = \mathcal{C}_{\text{FD},\text{ND}}\, \mathcal{C}_{\text{ND},\text{ND}}^{-1}\, \delta_{\text{ND}} + \epsilon,
\end{equation}
with matrix multiplication over modes, channels, and energies implicit.
The first term is the part of the FD flux uncertainty that is determined by the ND; it is the mean of $\delta_{\text{FD}}$ conditioned on $\delta_{\text{ND}}$ for Gaussian flux uncertainties.
The second term, $\epsilon$, is a random variable describing the residual FD flux uncertainties that are independent from ND flux uncertainties.
These residual uncertainties are distributed as $\epsilon^m_{\ell}(E_\nu) \sim \mathcal{N}(0, \mathcal{C}_{\text{FD}|\text{ND}})$, where
\begin{equation}
    \mathcal{C}_{\text{FD}|\text{ND}} = \mathcal{C}_{\text{FD},\text{FD}} - \mathcal{C}_{\text{FD},\text{ND}}\, \mathcal{C}_{\text{ND},\text{ND}}^{-1}\, \mathcal{C}_{\text{ND},\text{FD}}
\end{equation}
is the covariance matrix for FD flux uncertainties conditioned on the ND flux distribution.

Two random draws enter each realization of the flux systematic model.
First, to account for mismodeling of the ND flux, the ND flux $\Phi_{\text{ND}}$ in the model event rates Eqs.~\eqref{eq:ND_model_-} and~\eqref{eq:ND_model_+} is replaced, before training, by a random draw $\tilde{\Phi}_{\text{ND}}$ from the flux uncertainty model; this is the $\tilde{\Phi}_{\text{ND}}$ appearing in Eq.~\eqref{eq:deltaND}.
Second, to account for the residual FD flux uncertainty that cannot be learned from ND data, a random value of $\epsilon$ is drawn, and the unoscillated FD flux $\phi_{\text{FD}}$ in the model event rates Eqs.~\eqref{eq:FD_model_-} and~\eqref{eq:FD_model_+} is replaced by $\phi_{\text{FD}}\left[1 + \delta_{\text{FD}}\right]$ with $\delta_{\text{FD}}$ from Eq.~\eqref{eq:deltaFD}.
The simulated data, by contrast, are defined by the single fixed realization $a_j^\star$ throughout; only the model side is varied, as in a realistic analysis where the true flux is unknown but fixed.

This procedure differs slightly from the flux uncertainty sampling procedure used in Ref.~\cite{Hackett:2024dul}, in which noise was applied to model predictions only, but we find that both procedures give similar results.

After including flux nuisance parameters in ND training, the learned model is able to describe the data distribution both in principle and in practice as seen in Fig.~\ref{fig:confused_flux}.
In particular, variations in best-fit $\Delta m^2_{31}$ are at the $0.3\%$-level, consistent with other experimental systematics above and smaller than the $1\sigma$ distances between CIs.
The relatively large variations between flux draws seen in true toy-model cross section CIs are also removed by including $b_j$ nuisance parameters\footnote{In the true cross-section case, the flux nuisance parameters are set to the values used to generate the data rather than fitted.} as shown in the same figure.


\subsection{Network initialization}

Based on observations that ND training losses show no signs of saturation for either infinite or finite-statistics results, as well as more detailed studies of NN architecture variation in Ref.~\cite{Hackett:2024dul}, we assume that our NN ansatz is sufficiently expressive to capture the true toy-model cross sections.
Many different networks describe the ND data equally well, and which one training converges to depends on the random initialization. Retraining from different pseudorandom seeds therefore gives a sample of these equally good solutions. The spread among them does not capture the full approximation error of the NN ansatz, but it is a reasonable proxy for how much of the learned cross section is fixed by the data rather than by the training.

Oscillation parameter CI results using learned models with six different random NN initializations are shown in Fig.~\ref{fig:confused_init}.
Results are shown both for the idealized case of no experimental systematics, as well as for the case where finite statistics, detector resolution, and flux systematics are included.
Without experimental systematics, the effects of NN initialization are barely visible.
With experimental systematics, different NN initializations lead to visibly larger variations.
The fluctuations arising from NN initialization are still smaller than $1\sigma$ distances between CIs.
They are included in total uncertainty estimates below by using a different pseudorandom seed for NN initialization in each total systematic sample.


\subsection{Combined uncertainties}

\begin{figure}
    \centering
    \includegraphics[width=.9\linewidth]{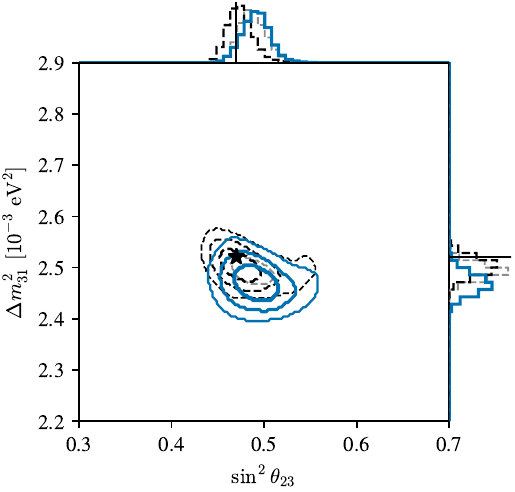}
    \includegraphics[width=.9\linewidth]{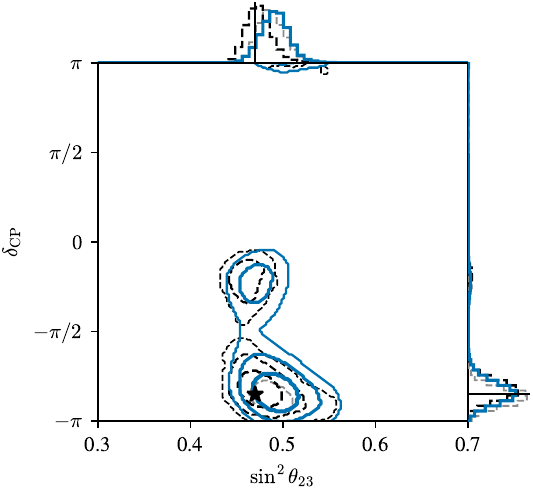}
    \includegraphics[width=.9\linewidth]{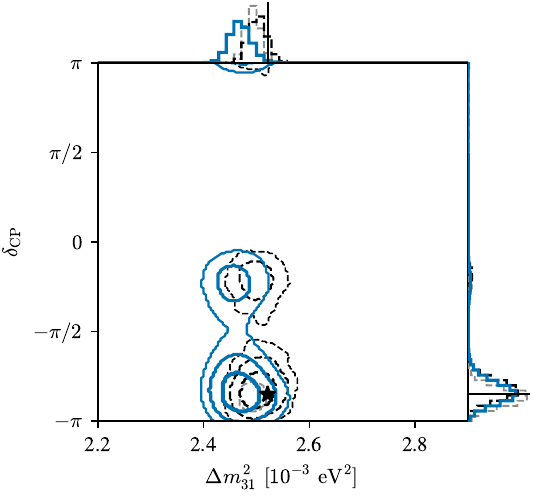} 
    \caption{Oscillation parameter CIs including systematic uncertainties as described in the main text for the true cross section (dashed black) and learned cross section (blue), and true cross section $1\sigma$ CIs without systematics (gray dashed).}
    \label{fig:combined_uncertainty}
\end{figure}

To estimate the combined effect of finite ND statistics, detector smearing, flux uncertainties, and NN initialization, we draw realizations of these systematics jointly. 
For each realization, we repeat the analysis using $N_{\rm boot}$ FD bootstrap samples. 
We then pool the fitted oscillation parameters from all $N_{\rm syst}$ realizations to construct the confidence intervals.

Let $\omega_{\star}^{(i,b)}$ be the best-fit oscillation parameters
obtained from FD bootstrap sample $b$ with systematic realization $i$.
We combine all of these fits into a single sample,
\begin{equation}
    \left\{\omega_{\star}^{(i,b)}
    \mid i=1,\ldots,N_{\rm syst},\;
    b=1,\ldots,N_{\rm boot}\right\},
\end{equation}
and construct confidence intervals from the pooled distribution.
This incorporates the variation across systematic realizations as well as the FD statistical variation within each realization.

Using $N_{\rm syst}=100$ systematic realizations and $N_{\rm boot}=4096$ FD bootstrap samples per realization, we obtain the confidence regions shown in Fig.~\ref{fig:combined_uncertainty}.
The median and the 16\% and 84\% quantiles of the pooled best-fit values are given for each oscillation parameter in Table~\ref{tab:param_std}.
For $\delta_{\rm CP}$, the $N_{\rm tot} = N_{\rm syst} N_{\rm boot}$ best-fit values are first unwrapped about their circular mean~\cite{Fisher} so that the quantiles account for the $2\pi$-periodicity.
The results show that the biases in best-fit oscillation parameters introduced by the learning procedure are smaller than or comparable to the corresponding $1\sigma$ statistical uncertainties.
Further, the $1\sigma$ widths obtained using the learned cross-section model are within 5--15\% of the corresponding CI widths obtained using the true toy-model cross section for all parameters, with the largest discrepancies arising in $\Delta m^2_{31}$.

\begin{table}[t]
\centering
\begin{tabular}{lccc}
\hline
Cross-section model & $\delta_{\rm CP} / \pi$ & $\sin^2\theta_{23}$ & $\Delta m^2_{31} / 10^{-3}$ \\ [2ex]
True cross section & $-0.863_{-50}^{+53}$ & $0.487_{-13}^{+16}$ & $2.493_{-14}^{+14}$ \\ [1ex]
Learned cross section & $-0.876_{-49}^{+52}$ & $0.491_{-12}^{+13}$ & $2.488_{-13}^{+13}$  \\ [1ex]
True w/ systematics & $-0.820_{-58}^{+79}$ & $0.473_{-12}^{+15}$ & $2.503_{-17}^{+17}$ \\ [1ex]
Learned w/ systematics & $-0.836_{-61}^{+80}$ & $0.491_{-15}^{+14}$ & $2.469_{-19}^{+21}$  \\
\hline
\end{tabular}
\caption{Comparison of oscillation parameter 68\% confidence intervals between analyses using true and learned cross-section models, with and without including systematic uncertainty estimates as described in the main text. The $2\pi$ periodicity of $\delta_{\rm CP}$ is accounted for using circular statistics~\cite{Fisher}.}
\label{tab:param_std}
\end{table}


\section{Conclusion}

We have extended the data-driven cross-section framework of Ref.~\cite{Hackett:2024dul} to include electron and muon final states, neutrino and antineutrino scattering, and both appearance and disappearance channels. 
Using a DUNE-inspired closure test, we trained neural representations of the nuclear structure functions on simulated ND event distributions and used the learned cross sections in a joint FD analysis of $\sin^2\theta_{23}$, $\Delta m^2_{31}$, and $\delta_{\rm CP}$.

In the idealized closure test, the oscillation constraints obtained with the learned cross section closely track those obtained with the true toy cross section.
We also find that finite ND statistics, detector energy resolution, and NN initialization have little effect on the results.
Flux uncertainties, however, affect the multi-channel oscillation analysis more strongly than in the previous single-channel analysis.
We mitigate their impact by fitting flux nuisance parameters alongside the neural cross-section model during ND training.
With this treatment, the oscillation constraints from the learned and true cross sections remain in close agreement even when flux uncertainties are included.

This closure test uses a known toy cross section and simplified detector and event samples, so it does not yet establish how well the method will perform on real experimental data. 
The next step is to test our approach with more realistic interaction models, detector responses, and backgrounds, and ultimately with ND measurements. 


\begin{acknowledgments}

We thank Minerba Betancourt, Laura Fields, Ian~Kotler, Kendall Mahn, Kevin McFarland, Bryan~Ramson, and DeMario Ross for useful discussions and helpful comments.

This document was prepared using the resources of the Fermi National Accelerator Laboratory (Fermilab), a U.S. Department of Energy, Office of Science, Office of High Energy Physics HEP User Facility. Fermilab is managed by Fermi Forward Discovery Group, LLC, acting under Contract No. 89243024CSC000002.
The work of K.T. is supported by DOE Grant KA2401045. The work of S.W.L.\ was supported by the National Science Foundation under Grant No.\ PHY-2544442 and URA Visiting Scholars Program Award \#2025-F-16. 
We thank the Institute for Nuclear Theory at the University of Washington for its kind hospitality and stimulating research environment. This research was supported in part by the INT's U.S. Department of Energy grant No.\ DE-FG02-00ER41132.
Numerical experiments and data analysis were performed using PyTorch~\cite{NEURIPS2019_9015}, NumPy~\cite{harris2020array}, SciPy~\cite{2020SciPy-NMeth}, pandas~\cite{jeff_reback_2020_3715232,mckinney-proc-scipy-2010}, gvar~\cite{peter_lepage_2020_4290884}, Mathematica~\cite{Mathematica}, and LHAPDF6~\cite{Buckley:2014ana}.
Figures were produced using matplotlib~\cite{Hunter:2007}.

{\bf AI Usage:} Claude by Anthropic was used as a coding assistant in this work. All physics choices, analysis design, and validation of results are the authors' own, and the authors take full responsibility for the content.

\end{acknowledgments}

\appendix


\section{Flux Covariance Matrix}
\label{app:flux_covariance}

\begin{figure*}
    \centering
    \includegraphics[width=\linewidth]{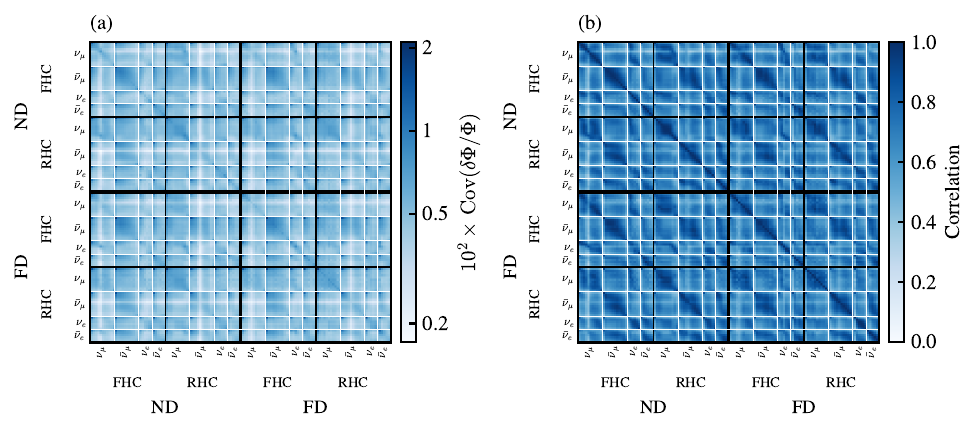}
    \caption{Left: Covariance matrix for the neutrino fluxes. The neutrino energy increases from top to bottom and from left to right. Right: Same as the left but showing the correlation matrix.}
    \label{fig:correlation}
\end{figure*}

The left panel of Fig.~\ref{fig:correlation} shows the flux covariance matrix of Refs.~\cite{DUNE:2020ypp,FieldsDune,IanKotler} that we use throughout. It is a fractional covariance, $\mathcal{C}_{ij} = \langle \delta\Phi_i\, \delta\Phi_j \rangle / (\Phi_i \Phi_j)$, so its diagonal gives the squared fractional uncertainty of each flux bin and the off-diagonal entries tell how a fluctuation in one bin goes with fluctuations in the others. The right panel shows the same information as a correlation matrix, $\mathcal{C}_{ij}/\sqrt{\mathcal{C}_{ii}\mathcal{C}_{jj}}$, which removes the overall size of the uncertainties and leaves only the correlation pattern, between $-1$ and $1$.

The bins are ordered by detector location (ND, then FD), then beam mode (FHC, then RHC), and then by flavor ($\nu_\mu$, $\bar\nu_\mu$, $\nu_e$, $\bar\nu_e$) with neutrino energy increasing within each block. Two features matter for this work. The ND and FD blocks of a given mode and flavor are strongly but not perfectly correlated, which is what allows the ND to constrain most, but not all, of the FD flux uncertainty (Sec.~\ref{subsec:fluxes}). And the correlations extend across flavors and modes, so a single flux realization moves all eight channels together.

\section{Training validation plots}
\label{app:training}

\begin{figure*}
    \centering
    \includegraphics[width=\linewidth]{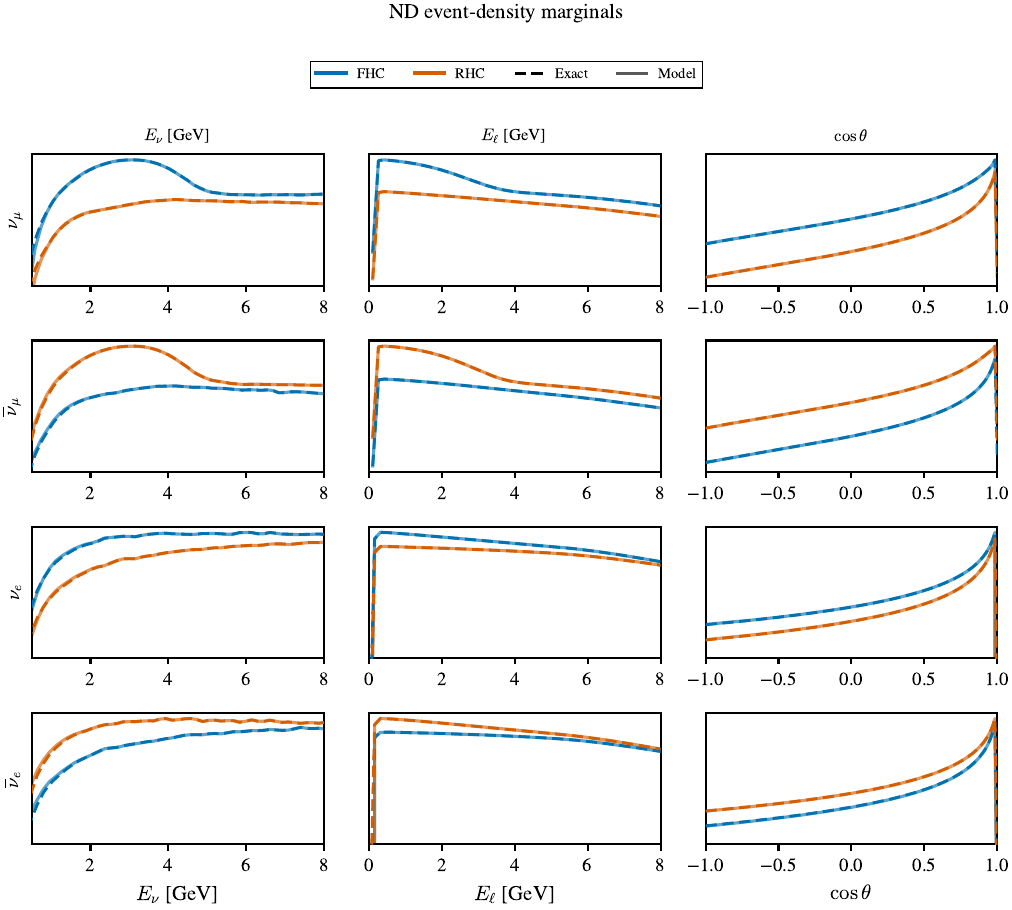}
    \caption{One-dimensional marginals of the true ND flux-weighted cross-section $\Phi_{\text{ND}}(E_\nu)\frac{d^2 \sigma^{(\nu A)}}{d\El d\costheta}(E_\nu)$ (dashed) are compared to their learned model counterparts (solid) for both FHC and RHC fluxes.}
    \label{fig:marginal_1D_ND}
\end{figure*}

\begin{figure*}
    \centering
    \includegraphics[width=\linewidth]{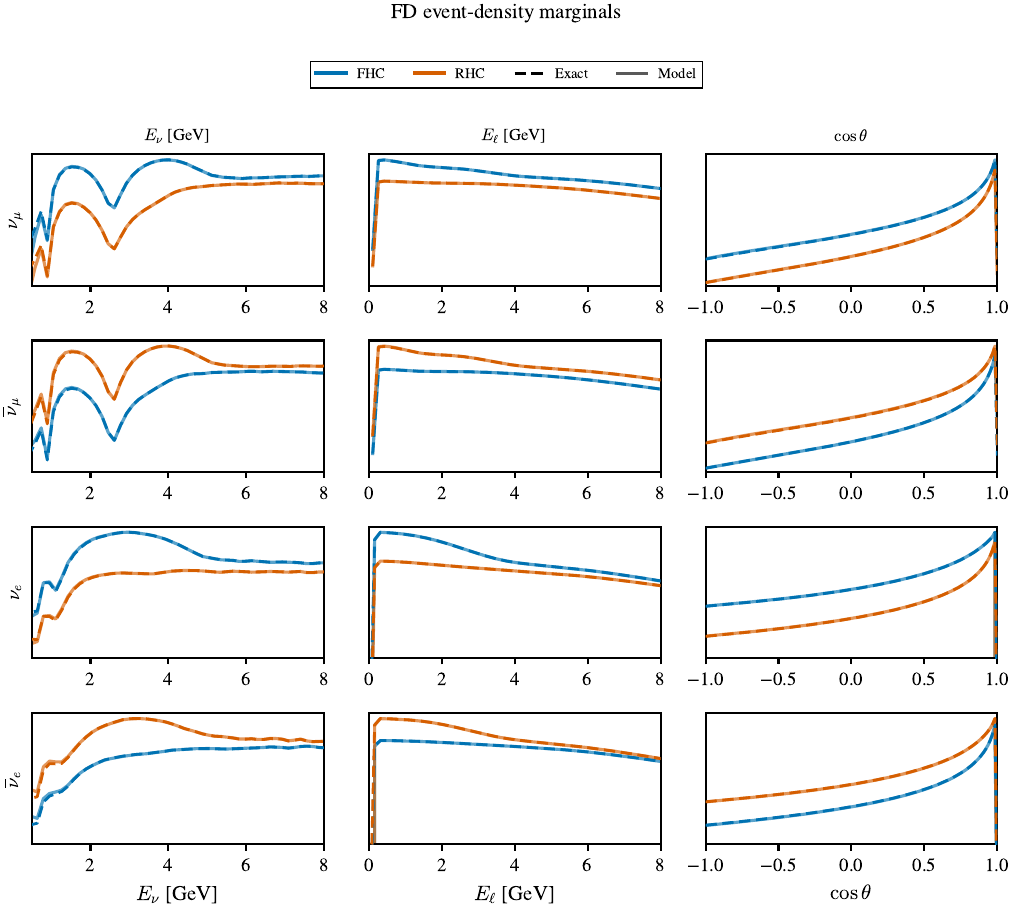}
    \caption{One-dimensional marginals of the true FD flux-weighted cross-section $\Phi_{\text{FD}}(E_\nu;\omega_{\rm true})\frac{d^2 \sigma^{(\nu A)}}{d\El d\costheta}(E_\nu)$ (dashed) are compared to their learned model counterparts (solid) for both FHC and RHC fluxes, both using the closure-test true oscillation parameters $\omega_{\rm true}$.}
    \label{fig:marginal_1D_FD}
\end{figure*}

\begin{figure}
    \centering
    \includegraphics[width=\linewidth]{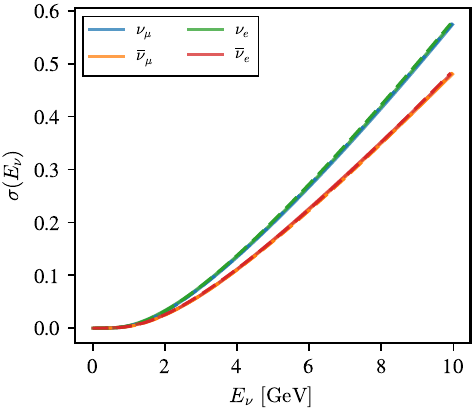}
    \caption{Total cross section $\sigma(E_\nu)$ for the four lepton channels, integrated over the $(v_1,v_2)$ region used in training, for the true toy model (dashed) and the learned model (solid). The vertical scale is in an arbitrary unit.}
    \label{fig:total_1d}
\end{figure}

The figures in this appendix show the trained cross-section model from a few more angles than in Sec.~\ref{sec:ND_results}. Figures~\ref{fig:marginal_1D_ND} and~\ref{fig:marginal_1D_FD} compare the one-dimensional marginals of the flux-weighted cross section, in $E_\nu$, $E_\ell$, $\cos\theta$, $v_1$ and $v_2$, for the true toy model (dashed) and the learned model (solid), using the ND and FD fluxes respectively. The FD case uses the oscillated flux at $\omega_{\rm true}$, so it directly tests the extrapolation from ND training to FD kinematics. The two agree over the full range of each variable, including the tails, which are several orders of magnitude below the peak. Figure~\ref{fig:total_1d} shows the total cross section $\sigma(E_\nu)$ for the four lepton channels, integrated over the $(v_1,v_2)$ region used in training. Here no normalization is applied to either curve, so the agreement is a check of the overall scale of the learned cross section, not just its shape.

\section{Systematic effects}
\label{app:systematics}

\begin{figure}
    \centering
    \includegraphics[width=\linewidth]{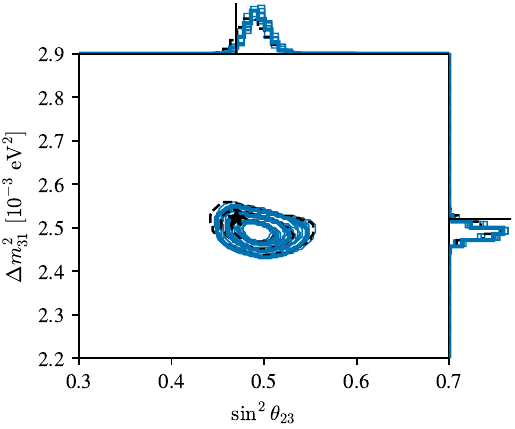}
    \includegraphics[width=\linewidth]{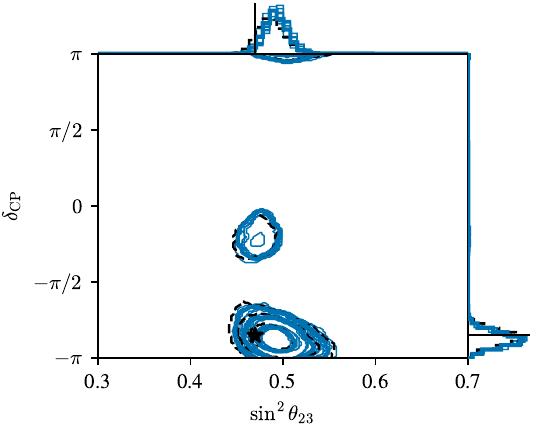}
    \includegraphics[width=\linewidth]{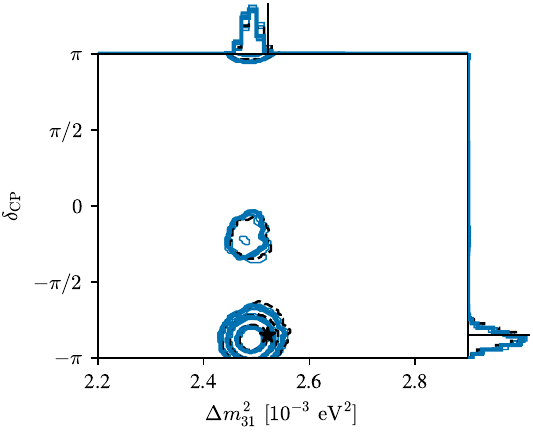}
    \caption{Confidence intervals as in Fig.~\ref{fig:confused_nosys} for six different draws of finite ND statistics, at the sample sizes of Table~\ref{tab:events_ND}, using the learned cross section (thin blue). The learned-model result with infinite ND statistics (thick blue) and the true cross-section result (black dashed) are shown for comparison.} 
    \label{fig:confused_stats}
\end{figure}

\begin{figure*}
    \centering
    \includegraphics[width=\linewidth]{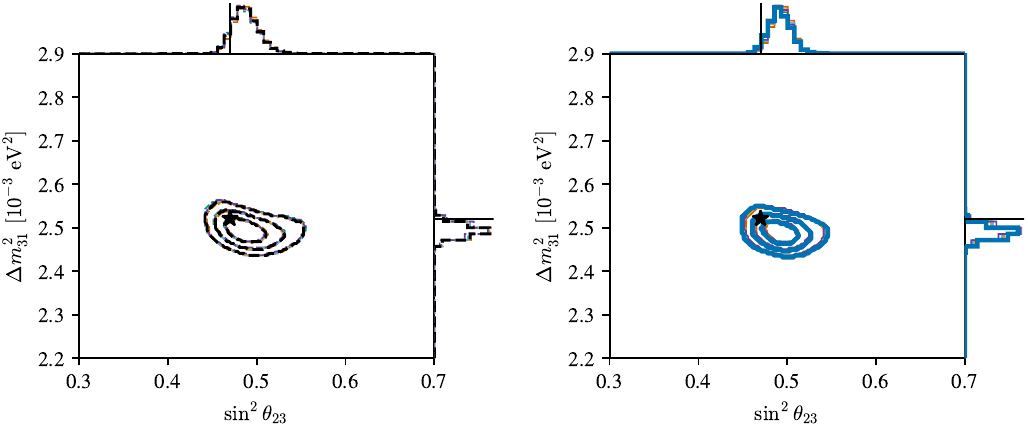}
    \includegraphics[width=\linewidth]{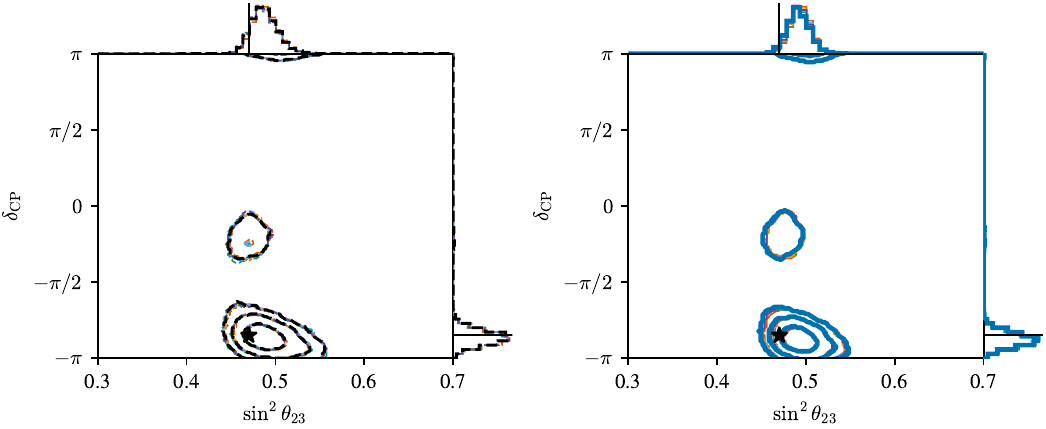}
    \includegraphics[width=\linewidth]{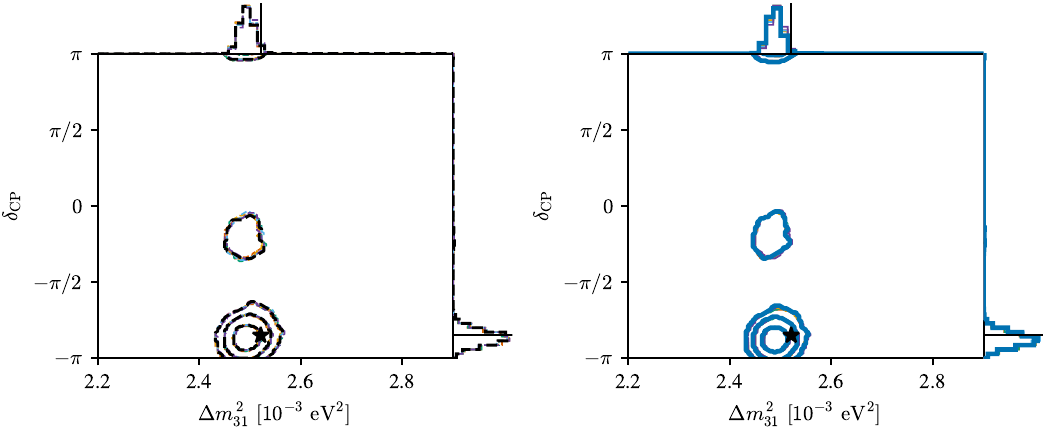}
    \caption{Left: confidence intervals for six different FD detector-smearing draws (dashed colored lines) using the true cross-section are compared with the corresponding results without detector smearing (dashed black lines). Right: learned cross-section results for the same six draws with ND detector smearing also included (thin colored lines) are compared with learned results without any detector smearing (thick blue lines).} 
    \label{fig:confused_smear}
\end{figure*}

\begin{figure*}
    \centering
    \includegraphics[width=\linewidth]{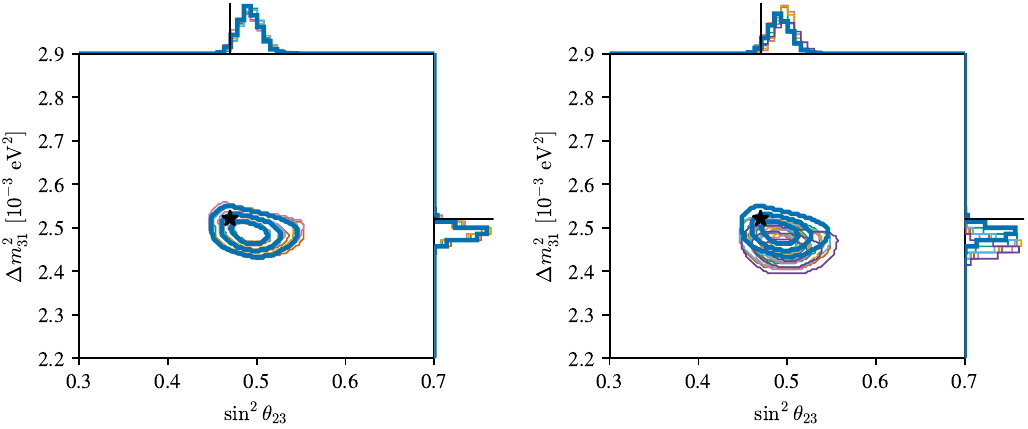}
    \includegraphics[width=\linewidth]{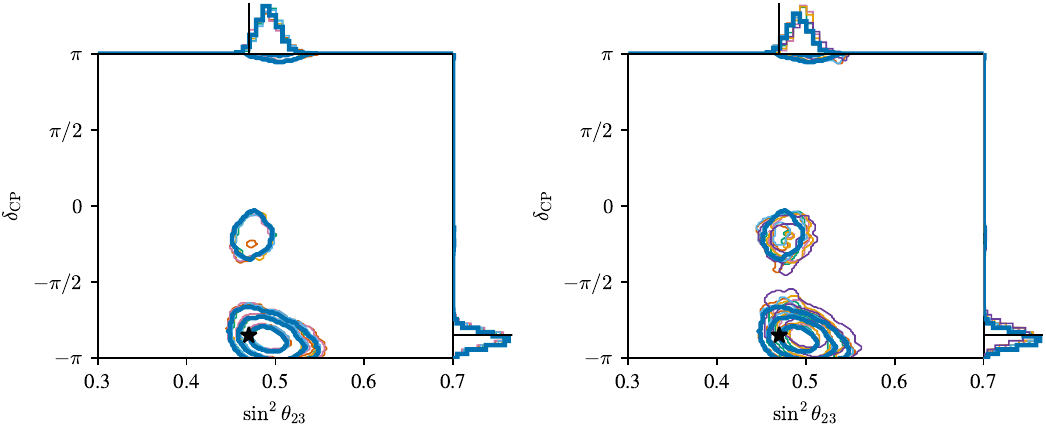} \\
    \includegraphics[width=\linewidth]{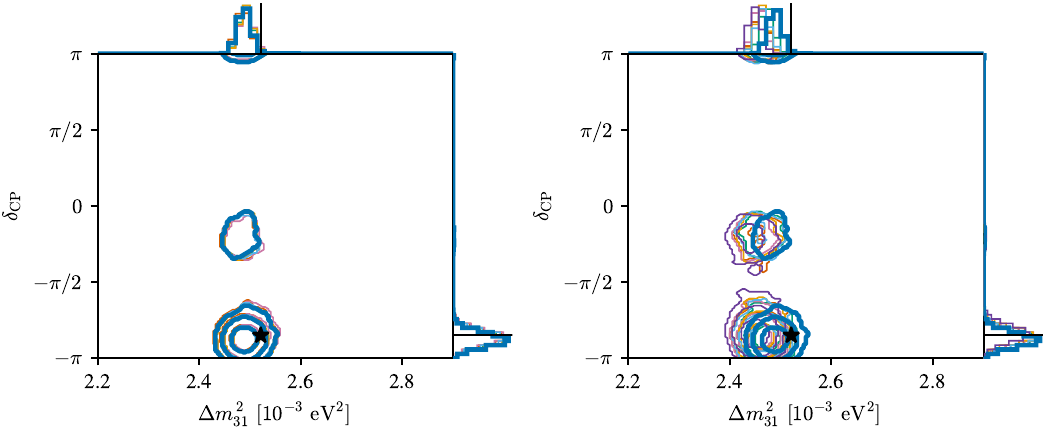}
    \caption{Confidence intervals as in Fig.~\ref{fig:confused_smear} for different random initializations of the NN parameters (i.e., different pseudo-random seeds). Left: without experimental systematic uncertainties. Right: with experimental finite-statistics, smearing, and flux systematics.}
    \label{fig:confused_init}
\end{figure*}

Section~\ref{sec:systematics} discusses the systematic effects one at a time but shows figures only for the flux uncertainties. The remaining ones are shown here. Figure~\ref{fig:confused_stats} shows six random draws of finite ND statistics for the learned cross section, compared with the infinite-statistics result. Figure~\ref{fig:confused_smear} shows detector smearing for the true cross section (left, dashed) and the learned cross section (right, solid), compared with the corresponding results without smearing (black dashed and thick blue). Figure~\ref{fig:confused_init} shows the variation with NN initialization, without (left) and with (right) the experimental systematics. In all three cases, the draws sit within the $1\sigma$ contour of the reference fit.

\section{Shape-only analysis}
\label{app:shape-only}

\begin{figure}
    \centering
    \includegraphics[width=.95\linewidth]{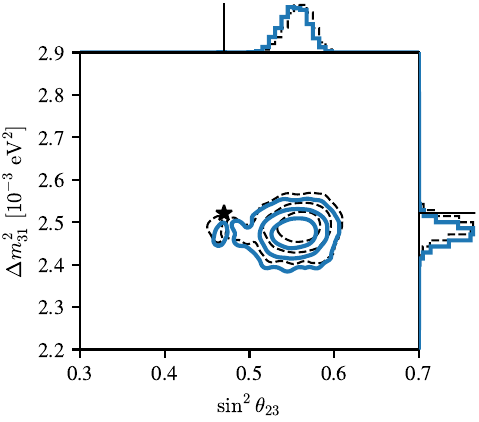}
    \includegraphics[width=.95\linewidth]{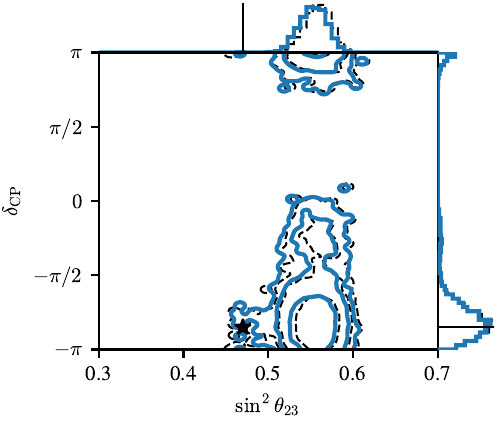}
    \includegraphics[width=.95\linewidth]{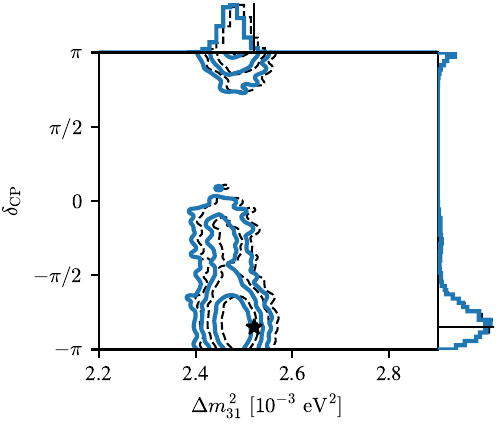}
    \caption{Shape-only FD analysis version of Fig.~\ref{fig:confused_nosys}. } 
    \label{fig:confused_nosys_shape}
\end{figure}

\begin{figure}
    \centering
    \includegraphics[width=.9\linewidth]{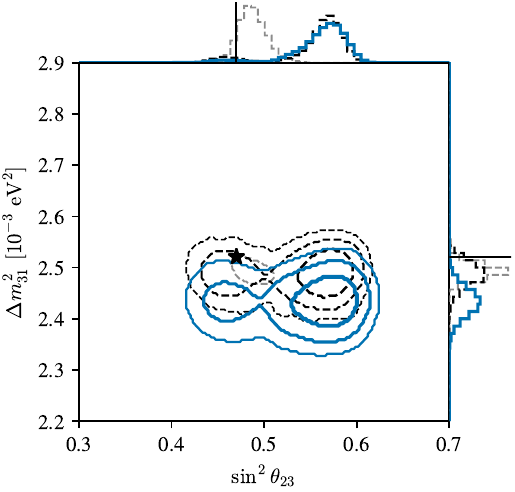}
    \includegraphics[width=.9\linewidth]{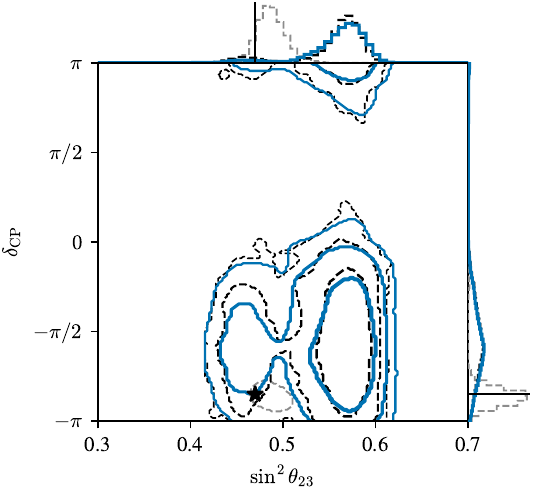}
    \includegraphics[width=.9\linewidth]{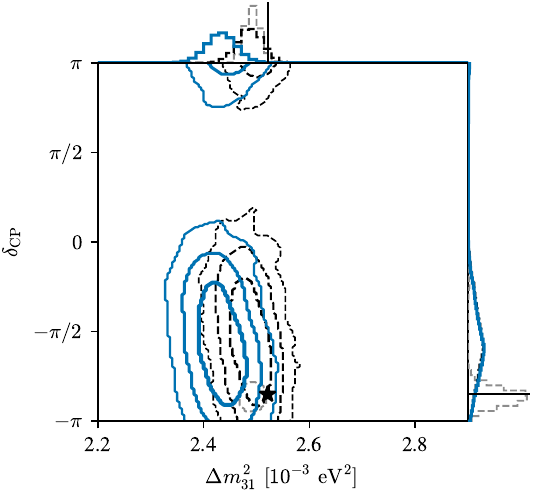} 
    \caption{Shape-only FD analysis version of Fig.~\ref{fig:combined_uncertainty}.}
    \label{fig:combined_uncertainty_shape}
\end{figure}

Reference~\cite{Hackett:2024dul} used a shape-only likelihood, in which each FD sample is normalized to its own predicted total so that only the distribution of events in $(E_\ell,\cos\theta)$ enters the fit. In this work, we include the normalization information per channel in our likelihood, in part because $\delta_\text{CP}$ is very sensitive to this information. For comparison, here we show the results obtained with the shape-only likelihood in Figs.~\ref{fig:confused_nosys_shape} and~\ref{fig:combined_uncertainty_shape}. Without the absolute rates and the ratios between samples, the worst degradation is seen in $\theta_{23}$: the wrong-octant solution is preferred for this FD sample, and the pooled result with systematics is bimodal in $\sin^2\theta_{23}$ with much wider contours. As expected, we also see a worsening of the precision of the $\delta_\text{CP}$ measurement. The learned cross section tracks the true one in this case too, so the comparison is a statement about the information in the data, not about the learning.

It is noteworthy that bias in $\Delta m^2_{31}$ and $\sin^2\theta_{23}$ is worse in the multi-channel analysis here than in the corresponding single-channel analysis in Ref.~\cite{Hackett:2024dul}.
This occurs because the more complicated ND training in the multi-channel case here results in shape and normalization mis-modeling effects that partially offset one another in a combined analysis and can more strongly bias a shape-only analysis.

\bibliographystyle{apsrev4-2}
\bibliography{refs}

\begin{thebibliography}{55}%
\makeatletter
\providecommand \@ifxundefined [1]{%
 \@ifx{#1\undefined}
}%
\providecommand \@ifnum [1]{%
 \ifnum #1\expandafter \@firstoftwo
 \else \expandafter \@secondoftwo
 \fi
}%
\providecommand \@ifx [1]{%
 \ifx #1\expandafter \@firstoftwo
 \else \expandafter \@secondoftwo
 \fi
}%
\providecommand \natexlab [1]{#1}%
\providecommand \enquote  [1]{``#1''}%
\providecommand \bibnamefont  [1]{#1}%
\providecommand \bibfnamefont [1]{#1}%
\providecommand \citenamefont [1]{#1}%
\providecommand \href@noop [0]{\@secondoftwo}%
\providecommand \href [0]{\begingroup \@sanitize@url \@href}%
\providecommand \@href[1]{\@@startlink{#1}\@@href}%
\providecommand \@@href[1]{\endgroup#1\@@endlink}%
\providecommand \@sanitize@url [0]{\catcode `\\12\catcode `\$12\catcode
  `\&12\catcode `\#12\catcode `\^12\catcode `\_12\catcode `\%12\relax}%
\providecommand \@@startlink[1]{}%
\providecommand \@@endlink[0]{}%
\providecommand \url  [0]{\begingroup\@sanitize@url \@url }%
\providecommand \@url [1]{\endgroup\@href {#1}{\urlprefix }}%
\providecommand \urlprefix  [0]{URL }%
\providecommand \Eprint [0]{\href }%
\providecommand \doibase [0]{https://doi.org/}%
\providecommand \selectlanguage [0]{\@gobble}%
\providecommand \bibinfo  [0]{\@secondoftwo}%
\providecommand \bibfield  [0]{\@secondoftwo}%
\providecommand \translation [1]{[#1]}%
\providecommand \BibitemOpen [0]{}%
\providecommand \bibitemStop [0]{}%
\providecommand \bibitemNoStop [0]{.\EOS\space}%
\providecommand \EOS [0]{\spacefactor3000\relax}%
\providecommand \BibitemShut  [1]{\csname bibitem#1\endcsname}%
\let\auto@bib@innerbib\@empty
\bibitem [{\citenamefont {Hackett}\ \emph {et~al.}(2025)\citenamefont
  {Hackett}, \citenamefont {Isaacson}, \citenamefont {Li}, \citenamefont
  {Tame-Narvaez},\ and\ \citenamefont {Wagman}}]{Hackett:2024dul}%
  \BibitemOpen
  \bibfield  {author} {\bibinfo {author} {\bibfnamefont {D.~C.}\ \bibnamefont
  {Hackett}}, \bibinfo {author} {\bibfnamefont {J.}~\bibnamefont {Isaacson}},
  \bibinfo {author} {\bibfnamefont {S.~W.}\ \bibnamefont {Li}}, \bibinfo
  {author} {\bibfnamefont {K.}~\bibnamefont {Tame-Narvaez}},\ and\ \bibinfo
  {author} {\bibfnamefont {M.~L.}\ \bibnamefont {Wagman}},\ }\href
  {https://doi.org/10.1103/bslk-gxr7} {\bibfield  {journal} {\bibinfo
  {journal} {Phys. Rev. D}\ }\textbf {\bibinfo {volume} {112}},\ \bibinfo
  {pages} {113005} (\bibinfo {year} {2025})},\ \Eprint
  {https://arxiv.org/abs/2412.16303} {arXiv:2412.16303 [hep-ph]} \BibitemShut
  {NoStop}%
\bibitem [{\citenamefont {An}\ \emph {et~al.}(2016)\citenamefont {An} \emph
  {et~al.}}]{JUNO:2015zny}%
  \BibitemOpen
  \bibfield  {author} {\bibinfo {author} {\bibfnamefont {F.}~\bibnamefont {An}}
  \emph {et~al.} (\bibinfo {collaboration} {JUNO}),\ }\href
  {https://doi.org/10.1088/0954-3899/43/3/030401} {\bibfield  {journal}
  {\bibinfo  {journal} {J. Phys. G}\ }\textbf {\bibinfo {volume} {43}},\
  \bibinfo {pages} {030401} (\bibinfo {year} {2016})},\ \Eprint
  {https://arxiv.org/abs/1507.05613} {arXiv:1507.05613 [physics.ins-det]}
  \BibitemShut {NoStop}%
\bibitem [{\citenamefont {Abe}\ \emph {et~al.}(2018)\citenamefont {Abe} \emph
  {et~al.}}]{Hyper-Kamiokande:2018ofw}%
  \BibitemOpen
  \bibfield  {author} {\bibinfo {author} {\bibfnamefont {K.}~\bibnamefont
  {Abe}} \emph {et~al.} (\bibinfo {collaboration} {Hyper-Kamiokande}),\
  }\href@noop {} {\  (\bibinfo {year} {2018})},\ \Eprint
  {https://arxiv.org/abs/1805.04163} {arXiv:1805.04163 [physics.ins-det]}
  \BibitemShut {NoStop}%
\bibitem [{\citenamefont {Abi}\ \emph {et~al.}(2020)\citenamefont {Abi} \emph
  {et~al.}}]{DUNE:2020ypp}%
  \BibitemOpen
  \bibfield  {author} {\bibinfo {author} {\bibfnamefont {B.}~\bibnamefont
  {Abi}} \emph {et~al.} (\bibinfo {collaboration} {DUNE}),\ }\href@noop {} {\
  (\bibinfo {year} {2020})},\ \Eprint {https://arxiv.org/abs/2002.03005}
  {arXiv:2002.03005 [hep-ex]} \BibitemShut {NoStop}%
\bibitem [{\citenamefont {Ruso}\ \emph {et~al.}(2025)\citenamefont {Ruso} \emph
  {et~al.}}]{Ruso:2022qes}%
  \BibitemOpen
  \bibfield  {author} {\bibinfo {author} {\bibfnamefont {L.~A.}\ \bibnamefont
  {Ruso}} \emph {et~al.},\ }\href {https://doi.org/10.1088/1361-6471/adae26}
  {\bibfield  {journal} {\bibinfo  {journal} {J. Phys. G}\ }\textbf {\bibinfo
  {volume} {52}},\ \bibinfo {pages} {043001} (\bibinfo {year} {2025})},\
  \Eprint {https://arxiv.org/abs/2203.09030} {arXiv:2203.09030 [hep-ph]}
  \BibitemShut {NoStop}%
\bibitem [{\citenamefont {de~Gouv{\^e}a}\ \emph {et~al.}(2022)\citenamefont
  {de~Gouv{\^e}a} \emph {et~al.}}]{deGouvea:2022gut}%
  \BibitemOpen
  \bibfield  {author} {\bibinfo {author} {\bibfnamefont {A.}~\bibnamefont
  {de~Gouv{\^e}a}} \emph {et~al.},\ }\href@noop {} {\  (\bibinfo {year}
  {2022})},\ \Eprint {https://arxiv.org/abs/2209.07983} {arXiv:2209.07983
  [hep-ph]} \BibitemShut {NoStop}%
\bibitem [{\citenamefont {Alvarez-Ruso}\ \emph {et~al.}(2018)\citenamefont
  {Alvarez-Ruso} \emph {et~al.}}]{NuSTEC:2017hzk}%
  \BibitemOpen
  \bibfield  {author} {\bibinfo {author} {\bibfnamefont {L.}~\bibnamefont
  {Alvarez-Ruso}} \emph {et~al.} (\bibinfo {collaboration} {NuSTEC}),\ }\href
  {https://doi.org/10.1016/j.ppnp.2018.01.006} {\bibfield  {journal} {\bibinfo
  {journal} {Prog. Part. Nucl. Phys.}\ }\textbf {\bibinfo {volume} {100}},\
  \bibinfo {pages} {1} (\bibinfo {year} {2018})},\ \Eprint
  {https://arxiv.org/abs/1706.03621} {arXiv:1706.03621 [hep-ph]} \BibitemShut
  {NoStop}%
\bibitem [{\citenamefont {Andreopoulos}\ \emph {et~al.}(2010)\citenamefont
  {Andreopoulos} \emph {et~al.}}]{Andreopoulos:2009rq}%
  \BibitemOpen
  \bibfield  {author} {\bibinfo {author} {\bibfnamefont {C.}~\bibnamefont
  {Andreopoulos}} \emph {et~al.},\ }\href
  {https://doi.org/10.1016/j.nima.2009.12.009} {\bibfield  {journal} {\bibinfo
  {journal} {Nucl. Instrum. Meth. A}\ }\textbf {\bibinfo {volume} {614}},\
  \bibinfo {pages} {87} (\bibinfo {year} {2010})},\ \Eprint
  {https://arxiv.org/abs/0905.2517} {arXiv:0905.2517 [hep-ph]} \BibitemShut
  {NoStop}%
\bibitem [{\citenamefont {Buss}\ \emph {et~al.}(2012)\citenamefont {Buss},
  \citenamefont {Gaitanos}, \citenamefont {Gallmeister}, \citenamefont {van
  Hees}, \citenamefont {Kaskulov}, \citenamefont {Lalakulich}, \citenamefont
  {Larionov}, \citenamefont {Leitner}, \citenamefont {Weil},\ and\
  \citenamefont {Mosel}}]{Buss:2011mx}%
  \BibitemOpen
  \bibfield  {author} {\bibinfo {author} {\bibfnamefont {O.}~\bibnamefont
  {Buss}}, \bibinfo {author} {\bibfnamefont {T.}~\bibnamefont {Gaitanos}},
  \bibinfo {author} {\bibfnamefont {K.}~\bibnamefont {Gallmeister}}, \bibinfo
  {author} {\bibfnamefont {H.}~\bibnamefont {van Hees}}, \bibinfo {author}
  {\bibfnamefont {M.}~\bibnamefont {Kaskulov}}, \bibinfo {author}
  {\bibfnamefont {O.}~\bibnamefont {Lalakulich}}, \bibinfo {author}
  {\bibfnamefont {A.~B.}\ \bibnamefont {Larionov}}, \bibinfo {author}
  {\bibfnamefont {T.}~\bibnamefont {Leitner}}, \bibinfo {author} {\bibfnamefont
  {J.}~\bibnamefont {Weil}},\ and\ \bibinfo {author} {\bibfnamefont
  {U.}~\bibnamefont {Mosel}},\ }\href
  {https://doi.org/10.1016/j.physrep.2011.12.001} {\bibfield  {journal}
  {\bibinfo  {journal} {Phys. Rept.}\ }\textbf {\bibinfo {volume} {512}},\
  \bibinfo {pages} {1} (\bibinfo {year} {2012})},\ \Eprint
  {https://arxiv.org/abs/1106.1344} {arXiv:1106.1344 [hep-ph]} \BibitemShut
  {NoStop}%
\bibitem [{\citenamefont {Golan}\ \emph {et~al.}(2012)\citenamefont {Golan},
  \citenamefont {Sobczyk},\ and\ \citenamefont {Zmuda}}]{Golan:2012rfa}%
  \BibitemOpen
  \bibfield  {author} {\bibinfo {author} {\bibfnamefont {T.}~\bibnamefont
  {Golan}}, \bibinfo {author} {\bibfnamefont {J.~T.}\ \bibnamefont {Sobczyk}},\
  and\ \bibinfo {author} {\bibfnamefont {J.}~\bibnamefont {Zmuda}},\ }\href
  {https://doi.org/10.1016/j.nuclphysbps.2012.09.136} {\bibfield  {journal}
  {\bibinfo  {journal} {Nucl. Phys. B Proc. Suppl.}\ }\textbf {\bibinfo
  {volume} {229-232}},\ \bibinfo {pages} {499} (\bibinfo {year}
  {2012})}\BibitemShut {NoStop}%
\bibitem [{\citenamefont {Hayato}\ and\ \citenamefont
  {Pickering}(2021)}]{Hayato:2021heg}%
  \BibitemOpen
  \bibfield  {author} {\bibinfo {author} {\bibfnamefont {Y.}~\bibnamefont
  {Hayato}}\ and\ \bibinfo {author} {\bibfnamefont {L.}~\bibnamefont
  {Pickering}},\ }\href {https://doi.org/10.1140/epjs/s11734-021-00287-7}
  {\bibfield  {journal} {\bibinfo  {journal} {Eur. Phys. J. ST}\ }\textbf
  {\bibinfo {volume} {230}},\ \bibinfo {pages} {4469} (\bibinfo {year}
  {2021})},\ \Eprint {https://arxiv.org/abs/2106.15809} {arXiv:2106.15809
  [hep-ph]} \BibitemShut {NoStop}%
\bibitem [{\citenamefont {Isaacson}\ \emph {et~al.}(2023)\citenamefont
  {Isaacson}, \citenamefont {Jay}, \citenamefont {Lovato}, \citenamefont
  {Machado},\ and\ \citenamefont {Rocco}}]{Isaacson:2022cwh}%
  \BibitemOpen
  \bibfield  {author} {\bibinfo {author} {\bibfnamefont {J.}~\bibnamefont
  {Isaacson}}, \bibinfo {author} {\bibfnamefont {W.~I.}\ \bibnamefont {Jay}},
  \bibinfo {author} {\bibfnamefont {A.}~\bibnamefont {Lovato}}, \bibinfo
  {author} {\bibfnamefont {P.~A.~N.}\ \bibnamefont {Machado}},\ and\ \bibinfo
  {author} {\bibfnamefont {N.}~\bibnamefont {Rocco}},\ }\href
  {https://doi.org/10.1103/PhysRevD.107.033007} {\bibfield  {journal} {\bibinfo
   {journal} {Phys. Rev. D}\ }\textbf {\bibinfo {volume} {107}},\ \bibinfo
  {pages} {033007} (\bibinfo {year} {2023})},\ \Eprint
  {https://arxiv.org/abs/2205.06378} {arXiv:2205.06378 [hep-ph]} \BibitemShut
  {NoStop}%
\bibitem [{\citenamefont {Acero}\ \emph {et~al.}(2020)\citenamefont {Acero}
  \emph {et~al.}}]{NOvA:2020rbg}%
  \BibitemOpen
  \bibfield  {author} {\bibinfo {author} {\bibfnamefont {M.~A.}\ \bibnamefont
  {Acero}} \emph {et~al.} (\bibinfo {collaboration} {NOvA}),\ }\href
  {https://doi.org/10.1140/epjc/s10052-020-08577-5} {\bibfield  {journal}
  {\bibinfo  {journal} {Eur. Phys. J. C}\ }\textbf {\bibinfo {volume} {80}},\
  \bibinfo {pages} {1119} (\bibinfo {year} {2020})},\ \Eprint
  {https://arxiv.org/abs/2006.08727} {arXiv:2006.08727 [hep-ex]} \BibitemShut
  {NoStop}%
\bibitem [{\citenamefont {Acero}\ \emph {et~al.}(2022)\citenamefont {Acero}
  \emph {et~al.}}]{NOvA:2021nfi}%
  \BibitemOpen
  \bibfield  {author} {\bibinfo {author} {\bibfnamefont {M.~A.}\ \bibnamefont
  {Acero}} \emph {et~al.} (\bibinfo {collaboration} {NOvA}),\ }\href
  {https://doi.org/10.1103/PhysRevD.106.032004} {\bibfield  {journal} {\bibinfo
   {journal} {Phys. Rev. D}\ }\textbf {\bibinfo {volume} {106}},\ \bibinfo
  {pages} {032004} (\bibinfo {year} {2022})},\ \Eprint
  {https://arxiv.org/abs/2108.08219} {arXiv:2108.08219 [hep-ex]} \BibitemShut
  {NoStop}%
\bibitem [{\citenamefont {Abe}\ \emph {et~al.}(2021)\citenamefont {Abe} \emph
  {et~al.}}]{T2K:2021xwb}%
  \BibitemOpen
  \bibfield  {author} {\bibinfo {author} {\bibfnamefont {K.}~\bibnamefont
  {Abe}} \emph {et~al.} (\bibinfo {collaboration} {T2K}),\ }\href
  {https://doi.org/10.1103/PhysRevD.103.112008} {\bibfield  {journal} {\bibinfo
   {journal} {Phys. Rev. D}\ }\textbf {\bibinfo {volume} {103}},\ \bibinfo
  {pages} {112008} (\bibinfo {year} {2021})},\ \Eprint
  {https://arxiv.org/abs/2101.03779} {arXiv:2101.03779 [hep-ex]} \BibitemShut
  {NoStop}%
\bibitem [{\citenamefont {Abe}\ \emph {et~al.}(2023)\citenamefont {Abe} \emph
  {et~al.}}]{T2K:2023smv}%
  \BibitemOpen
  \bibfield  {author} {\bibinfo {author} {\bibfnamefont {K.}~\bibnamefont
  {Abe}} \emph {et~al.} (\bibinfo {collaboration} {T2K}),\ }\href
  {https://doi.org/10.1140/epjc/s10052-023-11819-x} {\bibfield  {journal}
  {\bibinfo  {journal} {Eur. Phys. J. C}\ }\textbf {\bibinfo {volume} {83}},\
  \bibinfo {pages} {782} (\bibinfo {year} {2023})},\ \Eprint
  {https://arxiv.org/abs/2303.03222} {arXiv:2303.03222 [hep-ex]} \BibitemShut
  {NoStop}%
\bibitem [{\citenamefont {Stowell}\ \emph {et~al.}(2019)\citenamefont {Stowell}
  \emph {et~al.}}]{MINERvA:2019kfr}%
  \BibitemOpen
  \bibfield  {author} {\bibinfo {author} {\bibfnamefont {P.}~\bibnamefont
  {Stowell}} \emph {et~al.} (\bibinfo {collaboration} {MINERvA}),\ }\href
  {https://doi.org/10.1103/PhysRevD.100.072005} {\bibfield  {journal} {\bibinfo
   {journal} {Phys. Rev. D}\ }\textbf {\bibinfo {volume} {100}},\ \bibinfo
  {pages} {072005} (\bibinfo {year} {2019})},\ \Eprint
  {https://arxiv.org/abs/1903.01558} {arXiv:1903.01558 [hep-ex]} \BibitemShut
  {NoStop}%
\bibitem [{\citenamefont {Coyle}\ \emph {et~al.}(2022)\citenamefont {Coyle},
  \citenamefont {Li},\ and\ \citenamefont {Machado}}]{Coyle:2022bwa}%
  \BibitemOpen
  \bibfield  {author} {\bibinfo {author} {\bibfnamefont {N.~M.}\ \bibnamefont
  {Coyle}}, \bibinfo {author} {\bibfnamefont {S.~W.}\ \bibnamefont {Li}},\ and\
  \bibinfo {author} {\bibfnamefont {P.~A.~N.}\ \bibnamefont {Machado}},\ }\href
  {https://doi.org/10.1007/JHEP12(2022)166} {\bibfield  {journal} {\bibinfo
  {journal} {JHEP}\ }\textbf {\bibinfo {volume} {12}},\ \bibinfo {pages}
  {166}},\ \Eprint {https://arxiv.org/abs/2210.03753} {arXiv:2210.03753
  [hep-ph]} \BibitemShut {NoStop}%
\bibitem [{\citenamefont {Coyle}\ \emph {et~al.}(2025)\citenamefont {Coyle},
  \citenamefont {Li},\ and\ \citenamefont {Machado}}]{Coyle:2025xjk}%
  \BibitemOpen
  \bibfield  {author} {\bibinfo {author} {\bibfnamefont {N.~M.}\ \bibnamefont
  {Coyle}}, \bibinfo {author} {\bibfnamefont {S.~W.}\ \bibnamefont {Li}},\ and\
  \bibinfo {author} {\bibfnamefont {P.~A.~N.}\ \bibnamefont {Machado}},\ }\href
  {https://doi.org/10.1103/PhysRevD.111.093010} {\bibfield  {journal} {\bibinfo
   {journal} {Phys. Rev. D}\ }\textbf {\bibinfo {volume} {111}},\ \bibinfo
  {pages} {093010} (\bibinfo {year} {2025})},\ \Eprint
  {https://arxiv.org/abs/2502.19467} {arXiv:2502.19467 [hep-ph]} \BibitemShut
  {NoStop}%
\bibitem [{\citenamefont {Ball}\ \emph {et~al.}(2012)\citenamefont {Ball},
  \citenamefont {Bertone}, \citenamefont {Cerutti}, \citenamefont {Del~Debbio},
  \citenamefont {Forte}, \citenamefont {Guffanti}, \citenamefont {Latorre},
  \citenamefont {Rojo},\ and\ \citenamefont {Ubiali}}]{Ball:2011uy}%
  \BibitemOpen
  \bibfield  {author} {\bibinfo {author} {\bibfnamefont {R.~D.}\ \bibnamefont
  {Ball}}, \bibinfo {author} {\bibfnamefont {V.}~\bibnamefont {Bertone}},
  \bibinfo {author} {\bibfnamefont {F.}~\bibnamefont {Cerutti}}, \bibinfo
  {author} {\bibfnamefont {L.}~\bibnamefont {Del~Debbio}}, \bibinfo {author}
  {\bibfnamefont {S.}~\bibnamefont {Forte}}, \bibinfo {author} {\bibfnamefont
  {A.}~\bibnamefont {Guffanti}}, \bibinfo {author} {\bibfnamefont {J.~I.}\
  \bibnamefont {Latorre}}, \bibinfo {author} {\bibfnamefont {J.}~\bibnamefont
  {Rojo}},\ and\ \bibinfo {author} {\bibfnamefont {M.}~\bibnamefont {Ubiali}}
  (\bibinfo {collaboration} {NNPDF}),\ }\href
  {https://doi.org/10.1016/j.nuclphysb.2011.09.024} {\bibfield  {journal}
  {\bibinfo  {journal} {Nucl. Phys. B}\ }\textbf {\bibinfo {volume} {855}},\
  \bibinfo {pages} {153} (\bibinfo {year} {2012})},\ \Eprint
  {https://arxiv.org/abs/1107.2652} {arXiv:1107.2652 [hep-ph]} \BibitemShut
  {NoStop}%
\bibitem [{\citenamefont {Abdul~Khalek}\ \emph {et~al.}(2019)\citenamefont
  {Abdul~Khalek}, \citenamefont {Ethier},\ and\ \citenamefont
  {Rojo}}]{AbdulKhalek:2019mzd}%
  \BibitemOpen
  \bibfield  {author} {\bibinfo {author} {\bibfnamefont {R.}~\bibnamefont
  {Abdul~Khalek}}, \bibinfo {author} {\bibfnamefont {J.~J.}\ \bibnamefont
  {Ethier}},\ and\ \bibinfo {author} {\bibfnamefont {J.}~\bibnamefont {Rojo}}
  (\bibinfo {collaboration} {NNPDF}),\ }\href
  {https://doi.org/10.1140/epjc/s10052-019-6983-1} {\bibfield  {journal}
  {\bibinfo  {journal} {Eur. Phys. J. C}\ }\textbf {\bibinfo {volume} {79}},\
  \bibinfo {pages} {471} (\bibinfo {year} {2019})},\ \Eprint
  {https://arxiv.org/abs/1904.00018} {arXiv:1904.00018 [hep-ph]} \BibitemShut
  {NoStop}%
\bibitem [{\citenamefont {Forte}\ \emph {et~al.}(2002)\citenamefont {Forte},
  \citenamefont {Garrido}, \citenamefont {Latorre},\ and\ \citenamefont
  {Piccione}}]{Forte:2002fg}%
  \BibitemOpen
  \bibfield  {author} {\bibinfo {author} {\bibfnamefont {S.}~\bibnamefont
  {Forte}}, \bibinfo {author} {\bibfnamefont {L.}~\bibnamefont {Garrido}},
  \bibinfo {author} {\bibfnamefont {J.~I.}\ \bibnamefont {Latorre}},\ and\
  \bibinfo {author} {\bibfnamefont {A.}~\bibnamefont {Piccione}},\ }\href
  {https://doi.org/10.1088/1126-6708/2002/05/062} {\bibfield  {journal}
  {\bibinfo  {journal} {JHEP}\ }\textbf {\bibinfo {volume} {05}},\ \bibinfo
  {pages} {062}},\ \Eprint {https://arxiv.org/abs/hep-ph/0204232}
  {arXiv:hep-ph/0204232} \BibitemShut {NoStop}%
\bibitem [{\citenamefont {Sobczyk}\ \emph {et~al.}(2024)\citenamefont
  {Sobczyk}, \citenamefont {Rocco},\ and\ \citenamefont
  {Lovato}}]{Sobczyk:2024ajv}%
  \BibitemOpen
  \bibfield  {author} {\bibinfo {author} {\bibfnamefont {J.~E.}\ \bibnamefont
  {Sobczyk}}, \bibinfo {author} {\bibfnamefont {N.}~\bibnamefont {Rocco}},\
  and\ \bibinfo {author} {\bibfnamefont {A.}~\bibnamefont {Lovato}},\ }\href
  {https://doi.org/10.1016/j.physletb.2024.139142} {\bibfield  {journal}
  {\bibinfo  {journal} {Phys. Lett. B}\ }\textbf {\bibinfo {volume} {859}},\
  \bibinfo {pages} {139142} (\bibinfo {year} {2024})},\ \Eprint
  {https://arxiv.org/abs/2406.06292} {arXiv:2406.06292 [nucl-th]} \BibitemShut
  {NoStop}%
\bibitem [{\citenamefont {Bodek}\ \emph {et~al.}(2024)\citenamefont {Bodek},
  \citenamefont {Christy}, \citenamefont {Lin}, \citenamefont {Bulugean},
  \citenamefont {Delgado}, \citenamefont {Ankowski}, \citenamefont {Megias},\
  and\ \citenamefont {Tena~Vidal}}]{Bodek:2024mrp}%
  \BibitemOpen
  \bibfield  {author} {\bibinfo {author} {\bibfnamefont {A.}~\bibnamefont
  {Bodek}}, \bibinfo {author} {\bibfnamefont {M.~E.}\ \bibnamefont {Christy}},
  \bibinfo {author} {\bibfnamefont {Z.}~\bibnamefont {Lin}}, \bibinfo {author}
  {\bibfnamefont {G.-M.}\ \bibnamefont {Bulugean}}, \bibinfo {author}
  {\bibfnamefont {A.~M.}\ \bibnamefont {Delgado}}, \bibinfo {author}
  {\bibfnamefont {A.~M.}\ \bibnamefont {Ankowski}}, \bibinfo {author}
  {\bibfnamefont {G.~D.}\ \bibnamefont {Megias}},\ and\ \bibinfo {author}
  {\bibfnamefont {J.}~\bibnamefont {Tena~Vidal}},\ }\href@noop {} {\  (\bibinfo
  {year} {2024})},\ \Eprint {https://arxiv.org/abs/2409.10637}
  {arXiv:2409.10637 [hep-ex]} \BibitemShut {NoStop}%
\bibitem [{\citenamefont {Candido}\ \emph {et~al.}(2023)\citenamefont
  {Candido}, \citenamefont {Garcia}, \citenamefont {Magni}, \citenamefont
  {Rabemananjara}, \citenamefont {Rojo},\ and\ \citenamefont
  {Stegeman}}]{Candido:2023utz}%
  \BibitemOpen
  \bibfield  {author} {\bibinfo {author} {\bibfnamefont {A.}~\bibnamefont
  {Candido}}, \bibinfo {author} {\bibfnamefont {A.}~\bibnamefont {Garcia}},
  \bibinfo {author} {\bibfnamefont {G.}~\bibnamefont {Magni}}, \bibinfo
  {author} {\bibfnamefont {T.}~\bibnamefont {Rabemananjara}}, \bibinfo {author}
  {\bibfnamefont {J.}~\bibnamefont {Rojo}},\ and\ \bibinfo {author}
  {\bibfnamefont {R.}~\bibnamefont {Stegeman}},\ }\href
  {https://doi.org/10.1007/JHEP05(2023)149} {\bibfield  {journal} {\bibinfo
  {journal} {JHEP}\ }\textbf {\bibinfo {volume} {05}},\ \bibinfo {pages}
  {149}},\ \Eprint {https://arxiv.org/abs/2302.08527} {arXiv:2302.08527
  [hep-ph]} \BibitemShut {NoStop}%
\bibitem [{\citenamefont {Albright}\ and\ \citenamefont
  {Jarlskog}(1975)}]{Albright:1974ts}%
  \BibitemOpen
  \bibfield  {author} {\bibinfo {author} {\bibfnamefont {C.~H.}\ \bibnamefont
  {Albright}}\ and\ \bibinfo {author} {\bibfnamefont {C.}~\bibnamefont
  {Jarlskog}},\ }\href {https://doi.org/10.1016/0550-3213(75)90318-1}
  {\bibfield  {journal} {\bibinfo  {journal} {Nucl. Phys. B}\ }\textbf
  {\bibinfo {volume} {84}},\ \bibinfo {pages} {467} (\bibinfo {year}
  {1975})}\BibitemShut {NoStop}%
\bibitem [{\citenamefont {Paschos}\ and\ \citenamefont
  {Yu}(2002)}]{Paschos:2001np}%
  \BibitemOpen
  \bibfield  {author} {\bibinfo {author} {\bibfnamefont {E.~A.}\ \bibnamefont
  {Paschos}}\ and\ \bibinfo {author} {\bibfnamefont {J.~Y.}\ \bibnamefont
  {Yu}},\ }\href {https://doi.org/10.1103/PhysRevD.65.033002} {\bibfield
  {journal} {\bibinfo  {journal} {Phys. Rev. D}\ }\textbf {\bibinfo {volume}
  {65}},\ \bibinfo {pages} {033002} (\bibinfo {year} {2002})},\ \Eprint
  {https://arxiv.org/abs/hep-ph/0107261} {arXiv:hep-ph/0107261} \BibitemShut
  {NoStop}%
\bibitem [{\citenamefont {Kretzer}\ and\ \citenamefont
  {Reno}(2002)}]{Kretzer:2002fr}%
  \BibitemOpen
  \bibfield  {author} {\bibinfo {author} {\bibfnamefont {S.}~\bibnamefont
  {Kretzer}}\ and\ \bibinfo {author} {\bibfnamefont {M.~H.}\ \bibnamefont
  {Reno}},\ }\href {https://doi.org/10.1103/PhysRevD.66.113007} {\bibfield
  {journal} {\bibinfo  {journal} {Phys. Rev. D}\ }\textbf {\bibinfo {volume}
  {66}},\ \bibinfo {pages} {113007} (\bibinfo {year} {2002})},\ \Eprint
  {https://arxiv.org/abs/hep-ph/0208187} {arXiv:hep-ph/0208187} \BibitemShut
  {NoStop}%
\bibitem [{\citenamefont {Tomalak}\ \emph
  {et~al.}(2022{\natexlab{a}})\citenamefont {Tomalak}, \citenamefont {Chen},
  \citenamefont {Hill},\ and\ \citenamefont {McFarland}}]{Tomalak:2021hec}%
  \BibitemOpen
  \bibfield  {author} {\bibinfo {author} {\bibfnamefont {O.}~\bibnamefont
  {Tomalak}}, \bibinfo {author} {\bibfnamefont {Q.}~\bibnamefont {Chen}},
  \bibinfo {author} {\bibfnamefont {R.~J.}\ \bibnamefont {Hill}},\ and\
  \bibinfo {author} {\bibfnamefont {K.~S.}\ \bibnamefont {McFarland}},\ }\href
  {https://doi.org/10.1038/s41467-022-32974-x} {\bibfield  {journal} {\bibinfo
  {journal} {Nature Commun.}\ }\textbf {\bibinfo {volume} {13}},\ \bibinfo
  {pages} {5286} (\bibinfo {year} {2022}{\natexlab{a}})},\ \Eprint
  {https://arxiv.org/abs/2105.07939} {arXiv:2105.07939 [hep-ph]} \BibitemShut
  {NoStop}%
\bibitem [{\citenamefont {Tomalak}\ \emph
  {et~al.}(2022{\natexlab{b}})\citenamefont {Tomalak}, \citenamefont {Chen},
  \citenamefont {Hill}, \citenamefont {McFarland},\ and\ \citenamefont
  {Wret}}]{Tomalak:2022xup}%
  \BibitemOpen
  \bibfield  {author} {\bibinfo {author} {\bibfnamefont {O.}~\bibnamefont
  {Tomalak}}, \bibinfo {author} {\bibfnamefont {Q.}~\bibnamefont {Chen}},
  \bibinfo {author} {\bibfnamefont {R.~J.}\ \bibnamefont {Hill}}, \bibinfo
  {author} {\bibfnamefont {K.~S.}\ \bibnamefont {McFarland}},\ and\ \bibinfo
  {author} {\bibfnamefont {C.}~\bibnamefont {Wret}},\ }\href
  {https://doi.org/10.1103/PhysRevD.106.093006} {\bibfield  {journal} {\bibinfo
   {journal} {Phys. Rev. D}\ }\textbf {\bibinfo {volume} {106}},\ \bibinfo
  {pages} {093006} (\bibinfo {year} {2022}{\natexlab{b}})},\ \Eprint
  {https://arxiv.org/abs/2204.11379} {arXiv:2204.11379 [hep-ph]} \BibitemShut
  {NoStop}%
\bibitem [{\citenamefont {Afanasev}\ \emph {et~al.}(2024)\citenamefont
  {Afanasev} \emph {et~al.}}]{Afanasev:2023gev}%
  \BibitemOpen
  \bibfield  {author} {\bibinfo {author} {\bibfnamefont {A.}~\bibnamefont
  {Afanasev}} \emph {et~al.},\ }\href
  {https://doi.org/10.1140/epja/s10050-024-01281-y} {\bibfield  {journal}
  {\bibinfo  {journal} {Eur. Phys. J. A}\ }\textbf {\bibinfo {volume} {60}},\
  \bibinfo {pages} {91} (\bibinfo {year} {2024})},\ \Eprint
  {https://arxiv.org/abs/2306.14578} {arXiv:2306.14578 [hep-ph]} \BibitemShut
  {NoStop}%
\bibitem [{\citenamefont {Hewes}\ \emph {et~al.}(2021)\citenamefont {Hewes}
  \emph {et~al.}}]{DUNE:2021tad}%
  \BibitemOpen
  \bibfield  {author} {\bibinfo {author} {\bibfnamefont {V.}~\bibnamefont
  {Hewes}} \emph {et~al.} (\bibinfo {collaboration} {DUNE}),\ }\href
  {https://doi.org/10.3390/instruments5040031} {\bibfield  {journal} {\bibinfo
  {journal} {Instruments}\ }\textbf {\bibinfo {volume} {5}},\ \bibinfo {pages}
  {31} (\bibinfo {year} {2021})},\ \Eprint {https://arxiv.org/abs/2103.13910}
  {arXiv:2103.13910 [physics.ins-det]} \BibitemShut {NoStop}%
\bibitem [{\citenamefont {Fields}()}]{FieldsDune}%
  \BibitemOpen
  \bibfield  {author} {\bibinfo {author} {\bibfnamefont {L.}~\bibnamefont
  {Fields}},\ }\href@noop {} {\bibinfo {title} {{DUNE Fluxes}}},\ \bibinfo
  {note} {\url{https://glaucus.crc.nd.edu/DUNEFluxes/}}\BibitemShut {NoStop}%
\bibitem [{\citenamefont {Kotler}(2026)}]{IanKotler}%
  \BibitemOpen
  \bibfield  {author} {\bibinfo {author} {\bibfnamefont {I.}~\bibnamefont
  {Kotler}},\ }\href@noop {} {}\bibinfo {howpublished} {Private communication}
  (\bibinfo {year} {2026})\BibitemShut {NoStop}%
\bibitem [{\citenamefont {Esteban}\ \emph {et~al.}(2024)\citenamefont
  {Esteban}, \citenamefont {Gonzalez-Garcia}, \citenamefont {Maltoni},
  \citenamefont {Martinez-Soler}, \citenamefont {Pinheiro},\ and\ \citenamefont
  {Schwetz}}]{Esteban:2024eli}%
  \BibitemOpen
  \bibfield  {author} {\bibinfo {author} {\bibfnamefont {I.}~\bibnamefont
  {Esteban}}, \bibinfo {author} {\bibfnamefont {M.~C.}\ \bibnamefont
  {Gonzalez-Garcia}}, \bibinfo {author} {\bibfnamefont {M.}~\bibnamefont
  {Maltoni}}, \bibinfo {author} {\bibfnamefont {I.}~\bibnamefont
  {Martinez-Soler}}, \bibinfo {author} {\bibfnamefont {J.~P.}\ \bibnamefont
  {Pinheiro}},\ and\ \bibinfo {author} {\bibfnamefont {T.}~\bibnamefont
  {Schwetz}},\ }\href {https://doi.org/10.1007/JHEP12(2024)216} {\bibfield
  {journal} {\bibinfo  {journal} {JHEP}\ }\textbf {\bibinfo {volume} {12}},\
  \bibinfo {pages} {216}},\ \Eprint {https://arxiv.org/abs/2410.05380}
  {arXiv:2410.05380 [hep-ph]} \BibitemShut {NoStop}%
\bibitem [{NuF()}]{NuFIT:6.1}%
  \BibitemOpen
  \href {http://www.nu-fit.org/} {\bibinfo {title} {{NuFIT 6.1
  (2025)}}}\BibitemShut {NoStop}%
\bibitem [{\citenamefont {Denton}\ and\ \citenamefont
  {Parke}(2024)}]{Denton:2024pzc}%
  \BibitemOpen
  \bibfield  {author} {\bibinfo {author} {\bibfnamefont {P.~B.}\ \bibnamefont
  {Denton}}\ and\ \bibinfo {author} {\bibfnamefont {S.~J.}\ \bibnamefont
  {Parke}},\ }\href {https://doi.org/10.1103/PhysRevD.110.073005} {\bibfield
  {journal} {\bibinfo  {journal} {Phys. Rev. D}\ }\textbf {\bibinfo {volume}
  {110}},\ \bibinfo {pages} {073005} (\bibinfo {year} {2024})},\ \Eprint
  {https://arxiv.org/abs/2405.02400} {arXiv:2405.02400 [hep-ph]} \BibitemShut
  {NoStop}%
\bibitem [{\citenamefont {Bjorken}\ and\ \citenamefont
  {Paschos}(1969)}]{Bjorken:1969ja}%
  \BibitemOpen
  \bibfield  {author} {\bibinfo {author} {\bibfnamefont {J.~D.}\ \bibnamefont
  {Bjorken}}\ and\ \bibinfo {author} {\bibfnamefont {E.~A.}\ \bibnamefont
  {Paschos}},\ }\href {https://doi.org/10.1103/PhysRev.185.1975} {\bibfield
  {journal} {\bibinfo  {journal} {Phys. Rev.}\ }\textbf {\bibinfo {volume}
  {185}},\ \bibinfo {pages} {1975} (\bibinfo {year} {1969})}\BibitemShut
  {NoStop}%
\bibitem [{\citenamefont {Callan}\ and\ \citenamefont
  {Gross}(1969)}]{Callan:1969uq}%
  \BibitemOpen
  \bibfield  {author} {\bibinfo {author} {\bibfnamefont {C.~G.}\ \bibnamefont
  {Callan}, \bibfnamefont {Jr.}}\ and\ \bibinfo {author} {\bibfnamefont
  {D.~J.}\ \bibnamefont {Gross}},\ }\href
  {https://doi.org/10.1103/PhysRevLett.22.156} {\bibfield  {journal} {\bibinfo
  {journal} {Phys. Rev. Lett.}\ }\textbf {\bibinfo {volume} {22}},\ \bibinfo
  {pages} {156} (\bibinfo {year} {1969})}\BibitemShut {NoStop}%
\bibitem [{\citenamefont {Hou}\ \emph {et~al.}(2021)\citenamefont {Hou} \emph
  {et~al.}}]{Hou:2019efy}%
  \BibitemOpen
  \bibfield  {author} {\bibinfo {author} {\bibfnamefont {T.-J.}\ \bibnamefont
  {Hou}} \emph {et~al.},\ }\href {https://doi.org/10.1103/PhysRevD.103.014013}
  {\bibfield  {journal} {\bibinfo  {journal} {Phys. Rev. D}\ }\textbf {\bibinfo
  {volume} {103}},\ \bibinfo {pages} {014013} (\bibinfo {year} {2021})},\
  \Eprint {https://arxiv.org/abs/1912.10053} {arXiv:1912.10053 [hep-ph]}
  \BibitemShut {NoStop}%
\bibitem [{\citenamefont {Buckley}\ \emph {et~al.}(2015)\citenamefont
  {Buckley}, \citenamefont {Ferrando}, \citenamefont {Lloyd}, \citenamefont
  {Nordstr\"om}, \citenamefont {Page}, \citenamefont {R\"ufenacht},
  \citenamefont {Sch\"onherr},\ and\ \citenamefont {Watt}}]{Buckley:2014ana}%
  \BibitemOpen
  \bibfield  {author} {\bibinfo {author} {\bibfnamefont {A.}~\bibnamefont
  {Buckley}}, \bibinfo {author} {\bibfnamefont {J.}~\bibnamefont {Ferrando}},
  \bibinfo {author} {\bibfnamefont {S.}~\bibnamefont {Lloyd}}, \bibinfo
  {author} {\bibfnamefont {K.}~\bibnamefont {Nordstr\"om}}, \bibinfo {author}
  {\bibfnamefont {B.}~\bibnamefont {Page}}, \bibinfo {author} {\bibfnamefont
  {M.}~\bibnamefont {R\"ufenacht}}, \bibinfo {author} {\bibfnamefont
  {M.}~\bibnamefont {Sch\"onherr}},\ and\ \bibinfo {author} {\bibfnamefont
  {G.}~\bibnamefont {Watt}},\ }\href
  {https://doi.org/10.1140/epjc/s10052-015-3318-8} {\bibfield  {journal}
  {\bibinfo  {journal} {Eur. Phys. J. C}\ }\textbf {\bibinfo {volume} {75}},\
  \bibinfo {pages} {132} (\bibinfo {year} {2015})},\ \Eprint
  {https://arxiv.org/abs/1412.7420} {arXiv:1412.7420 [hep-ph]} \BibitemShut
  {NoStop}%
\bibitem [{\citenamefont {Martin}\ \emph {et~al.}(2009)\citenamefont {Martin},
  \citenamefont {Stirling}, \citenamefont {Thorne},\ and\ \citenamefont
  {Watt}}]{Martin:2009iq}%
  \BibitemOpen
  \bibfield  {author} {\bibinfo {author} {\bibfnamefont {A.~D.}\ \bibnamefont
  {Martin}}, \bibinfo {author} {\bibfnamefont {W.~J.}\ \bibnamefont
  {Stirling}}, \bibinfo {author} {\bibfnamefont {R.~S.}\ \bibnamefont
  {Thorne}},\ and\ \bibinfo {author} {\bibfnamefont {G.}~\bibnamefont {Watt}},\
  }\href {https://doi.org/10.1140/epjc/s10052-009-1072-5} {\bibfield  {journal}
  {\bibinfo  {journal} {Eur. Phys. J. C}\ }\textbf {\bibinfo {volume} {63}},\
  \bibinfo {pages} {189} (\bibinfo {year} {2009})},\ \Eprint
  {https://arxiv.org/abs/0901.0002} {arXiv:0901.0002 [hep-ph]} \BibitemShut
  {NoStop}%
\bibitem [{\citenamefont {Kingma}\ and\ \citenamefont
  {Ba}(2014)}]{Kingma:2014vow}%
  \BibitemOpen
  \bibfield  {author} {\bibinfo {author} {\bibfnamefont {D.~P.}\ \bibnamefont
  {Kingma}}\ and\ \bibinfo {author} {\bibfnamefont {J.}~\bibnamefont {Ba}}\
  }(\bibinfo {year} {2014})\ \Eprint {https://arxiv.org/abs/1412.6980}
  {arXiv:1412.6980 [cs.LG]} \BibitemShut {NoStop}%
\bibitem [{\citenamefont {Efron}(1982)}]{Efron:1982}%
  \BibitemOpen
  \bibfield  {author} {\bibinfo {author} {\bibfnamefont {B.}~\bibnamefont
  {Efron}},\ }\href
  {http://gso.gbv.de/DB=2.1/CMD?ACT=SRCHA&SRT=YOP&IKT=1016&TRM=ppn+017474957&sourceid=fbw_bibsonomy}
  {\emph {\bibinfo {title} {The Jackknife, the bootstrap and other resampling
  plans}}},\ \bibinfo {series} {Regional Conference Series in applied
  mathematics}\ No.~\bibinfo {number} {38}\ (\bibinfo  {publisher} {Society for
  Industrial and Applied Mathematics},\ \bibinfo {address} {Philadelphia,
  Pa.},\ \bibinfo {year} {1982})\BibitemShut {NoStop}%
\bibitem [{\citenamefont {DiCiccio}\ and\ \citenamefont
  {Efron}(1996)}]{DiCiccio:1996}%
  \BibitemOpen
  \bibfield  {author} {\bibinfo {author} {\bibfnamefont {T.~J.}\ \bibnamefont
  {DiCiccio}}\ and\ \bibinfo {author} {\bibfnamefont {B.}~\bibnamefont
  {Efron}},\ }\href {https://doi.org/10.1214/ss/1032280214} {\bibfield
  {journal} {\bibinfo  {journal} {Statistical Science}\ }\textbf {\bibinfo
  {volume} {11}},\ \bibinfo {pages} {189 } (\bibinfo {year}
  {1996})}\BibitemShut {NoStop}%
\bibitem [{\citenamefont {Davison}\ and\ \citenamefont
  {Hinkley}(1997)}]{Davison:1997}%
  \BibitemOpen
  \bibfield  {author} {\bibinfo {author} {\bibfnamefont {A.~C.}\ \bibnamefont
  {Davison}}\ and\ \bibinfo {author} {\bibfnamefont {D.~V.}\ \bibnamefont
  {Hinkley}},\ }\bibinfo {title} {The basic bootstraps},\ in\ \href
  {https://doi.org/10.1017/CBO9780511802843.003} {\emph {\bibinfo {booktitle}
  {Bootstrap Methods and their Application}}},\ \bibinfo {series and number}
  {Cambridge Series in Statistical and Probabilistic Mathematics}\ (\bibinfo
  {publisher} {Cambridge University Press},\ \bibinfo {year} {1997})\ p.\
  \bibinfo {pages} {11–69}\BibitemShut {NoStop}%
\bibitem [{\citenamefont {Paszke}\ \emph {et~al.}(2019)\citenamefont {Paszke}
  \emph {et~al.}}]{NEURIPS2019_9015}%
  \BibitemOpen
  \bibfield  {author} {\bibinfo {author} {\bibfnamefont {A.}~\bibnamefont
  {Paszke}} \emph {et~al.},\ }in\ \href
  {http://papers.neurips.cc/paper/9015-pytorch-an-imperative-style-high-performance\\-deep-learning-library.pdf}
  {\emph {\bibinfo {booktitle} {Advances in Neural Information Processing
  Systems 32}}},\ \bibinfo {editor} {edited by\ \bibinfo {editor}
  {\bibfnamefont {H.}~\bibnamefont {Wallach}}, \bibinfo {editor} {\bibfnamefont
  {H.}~\bibnamefont {Larochelle}}, \bibinfo {editor} {\bibfnamefont
  {A.}~\bibnamefont {Beygelzimer}}, \bibinfo {editor} {\bibfnamefont
  {F.}~\bibnamefont {d\textquotesingle Alch\'{e}-Buc}}, \bibinfo {editor}
  {\bibfnamefont {E.}~\bibnamefont {Fox}},\ and\ \bibinfo {editor}
  {\bibfnamefont {R.}~\bibnamefont {Garnett}}}\ (\bibinfo  {publisher} {Curran
  Associates, Inc.},\ \bibinfo {year} {2019})\ pp.\ \bibinfo {pages}
  {8024--8035}\BibitemShut {NoStop}%
\bibitem [{\citenamefont {Fisher}(1993)}]{Fisher}%
  \BibitemOpen
  \bibfield  {author} {\bibinfo {author} {\bibfnamefont {N.~I.}\ \bibnamefont
  {Fisher}},\ }\href {https://doi.org/10.1017/CBO9780511564345} {\emph
  {\bibinfo {title} {Statistical Analysis of Circular Data}}}\ (\bibinfo
  {publisher} {Cambridge University Press},\ \bibinfo {address} {Cambridge},\
  \bibinfo {year} {1993})\BibitemShut {NoStop}%
\bibitem [{\citenamefont {Harris}\ \emph {et~al.}(2020)\citenamefont {Harris},
  \citenamefont {Millman}, \citenamefont {van~der Walt}, \citenamefont
  {Gommers}, \citenamefont {Virtanen}, \citenamefont {Cournapeau},
  \citenamefont {Wieser}, \citenamefont {Taylor}, \citenamefont {Berg},
  \citenamefont {Smith}, \citenamefont {Kern}, \citenamefont {Picus},
  \citenamefont {Hoyer}, \citenamefont {van Kerkwijk}, \citenamefont {Brett},
  \citenamefont {Haldane}, \citenamefont {del R{\'{i}}o}, \citenamefont
  {Wiebe}, \citenamefont {Peterson}, \citenamefont {G{\'{e}}rard-Marchant},
  \citenamefont {Sheppard}, \citenamefont {Reddy}, \citenamefont {Weckesser},
  \citenamefont {Abbasi}, \citenamefont {Gohlke},\ and\ \citenamefont
  {Oliphant}}]{harris2020array}%
  \BibitemOpen
  \bibfield  {author} {\bibinfo {author} {\bibfnamefont {C.~R.}\ \bibnamefont
  {Harris}}, \bibinfo {author} {\bibfnamefont {K.~J.}\ \bibnamefont {Millman}},
  \bibinfo {author} {\bibfnamefont {S.~J.}\ \bibnamefont {van~der Walt}},
  \bibinfo {author} {\bibfnamefont {R.}~\bibnamefont {Gommers}}, \bibinfo
  {author} {\bibfnamefont {P.}~\bibnamefont {Virtanen}}, \bibinfo {author}
  {\bibfnamefont {D.}~\bibnamefont {Cournapeau}}, \bibinfo {author}
  {\bibfnamefont {E.}~\bibnamefont {Wieser}}, \bibinfo {author} {\bibfnamefont
  {J.}~\bibnamefont {Taylor}}, \bibinfo {author} {\bibfnamefont
  {S.}~\bibnamefont {Berg}}, \bibinfo {author} {\bibfnamefont {N.~J.}\
  \bibnamefont {Smith}}, \bibinfo {author} {\bibfnamefont {R.}~\bibnamefont
  {Kern}}, \bibinfo {author} {\bibfnamefont {M.}~\bibnamefont {Picus}},
  \bibinfo {author} {\bibfnamefont {S.}~\bibnamefont {Hoyer}}, \bibinfo
  {author} {\bibfnamefont {M.~H.}\ \bibnamefont {van Kerkwijk}}, \bibinfo
  {author} {\bibfnamefont {M.}~\bibnamefont {Brett}}, \bibinfo {author}
  {\bibfnamefont {A.}~\bibnamefont {Haldane}}, \bibinfo {author} {\bibfnamefont
  {J.~F.}\ \bibnamefont {del R{\'{i}}o}}, \bibinfo {author} {\bibfnamefont
  {M.}~\bibnamefont {Wiebe}}, \bibinfo {author} {\bibfnamefont
  {P.}~\bibnamefont {Peterson}}, \bibinfo {author} {\bibfnamefont
  {P.}~\bibnamefont {G{\'{e}}rard-Marchant}}, \bibinfo {author} {\bibfnamefont
  {K.}~\bibnamefont {Sheppard}}, \bibinfo {author} {\bibfnamefont
  {T.}~\bibnamefont {Reddy}}, \bibinfo {author} {\bibfnamefont
  {W.}~\bibnamefont {Weckesser}}, \bibinfo {author} {\bibfnamefont
  {H.}~\bibnamefont {Abbasi}}, \bibinfo {author} {\bibfnamefont
  {C.}~\bibnamefont {Gohlke}},\ and\ \bibinfo {author} {\bibfnamefont {T.~E.}\
  \bibnamefont {Oliphant}},\ }\href {https://doi.org/10.1038/s41586-020-2649-2}
  {\bibfield  {journal} {\bibinfo  {journal} {Nature}\ }\textbf {\bibinfo
  {volume} {585}},\ \bibinfo {pages} {357} (\bibinfo {year}
  {2020})}\BibitemShut {NoStop}%
\bibitem [{\citenamefont {Virtanen}\ \emph {et~al.}(2020)\citenamefont
  {Virtanen}, \citenamefont {Gommers}, \citenamefont {Oliphant}, \citenamefont
  {Haberland}, \citenamefont {Reddy}, \citenamefont {Cournapeau}, \citenamefont
  {Burovski}, \citenamefont {Peterson}, \citenamefont {Weckesser},
  \citenamefont {Bright}, \citenamefont {{van der Walt}}, \citenamefont
  {Brett}, \citenamefont {Wilson}, \citenamefont {Millman}, \citenamefont
  {Mayorov}, \citenamefont {Nelson}, \citenamefont {Jones}, \citenamefont
  {Kern}, \citenamefont {Larson}, \citenamefont {Carey}, \citenamefont {Polat},
  \citenamefont {Feng}, \citenamefont {Moore}, \citenamefont {{VanderPlas}},
  \citenamefont {Laxalde}, \citenamefont {Perktold}, \citenamefont {Cimrman},
  \citenamefont {Henriksen}, \citenamefont {Quintero}, \citenamefont {Harris},
  \citenamefont {Archibald}, \citenamefont {Ribeiro}, \citenamefont
  {Pedregosa}, \citenamefont {{van Mulbregt}},\ and\ \citenamefont {{SciPy 1.0
  Contributors}}}]{2020SciPy-NMeth}%
  \BibitemOpen
  \bibfield  {author} {\bibinfo {author} {\bibfnamefont {P.}~\bibnamefont
  {Virtanen}}, \bibinfo {author} {\bibfnamefont {R.}~\bibnamefont {Gommers}},
  \bibinfo {author} {\bibfnamefont {T.~E.}\ \bibnamefont {Oliphant}}, \bibinfo
  {author} {\bibfnamefont {M.}~\bibnamefont {Haberland}}, \bibinfo {author}
  {\bibfnamefont {T.}~\bibnamefont {Reddy}}, \bibinfo {author} {\bibfnamefont
  {D.}~\bibnamefont {Cournapeau}}, \bibinfo {author} {\bibfnamefont
  {E.}~\bibnamefont {Burovski}}, \bibinfo {author} {\bibfnamefont
  {P.}~\bibnamefont {Peterson}}, \bibinfo {author} {\bibfnamefont
  {W.}~\bibnamefont {Weckesser}}, \bibinfo {author} {\bibfnamefont
  {J.}~\bibnamefont {Bright}}, \bibinfo {author} {\bibfnamefont {S.~J.}\
  \bibnamefont {{van der Walt}}}, \bibinfo {author} {\bibfnamefont
  {M.}~\bibnamefont {Brett}}, \bibinfo {author} {\bibfnamefont
  {J.}~\bibnamefont {Wilson}}, \bibinfo {author} {\bibfnamefont {K.~J.}\
  \bibnamefont {Millman}}, \bibinfo {author} {\bibfnamefont {N.}~\bibnamefont
  {Mayorov}}, \bibinfo {author} {\bibfnamefont {A.~R.~J.}\ \bibnamefont
  {Nelson}}, \bibinfo {author} {\bibfnamefont {E.}~\bibnamefont {Jones}},
  \bibinfo {author} {\bibfnamefont {R.}~\bibnamefont {Kern}}, \bibinfo {author}
  {\bibfnamefont {E.}~\bibnamefont {Larson}}, \bibinfo {author} {\bibfnamefont
  {C.~J.}\ \bibnamefont {Carey}}, \bibinfo {author} {\bibfnamefont
  {{\.I}.}~\bibnamefont {Polat}}, \bibinfo {author} {\bibfnamefont
  {Y.}~\bibnamefont {Feng}}, \bibinfo {author} {\bibfnamefont {E.~W.}\
  \bibnamefont {Moore}}, \bibinfo {author} {\bibfnamefont {J.}~\bibnamefont
  {{VanderPlas}}}, \bibinfo {author} {\bibfnamefont {D.}~\bibnamefont
  {Laxalde}}, \bibinfo {author} {\bibfnamefont {J.}~\bibnamefont {Perktold}},
  \bibinfo {author} {\bibfnamefont {R.}~\bibnamefont {Cimrman}}, \bibinfo
  {author} {\bibfnamefont {I.}~\bibnamefont {Henriksen}}, \bibinfo {author}
  {\bibfnamefont {E.~A.}\ \bibnamefont {Quintero}}, \bibinfo {author}
  {\bibfnamefont {C.~R.}\ \bibnamefont {Harris}}, \bibinfo {author}
  {\bibfnamefont {A.~M.}\ \bibnamefont {Archibald}}, \bibinfo {author}
  {\bibfnamefont {A.~H.}\ \bibnamefont {Ribeiro}}, \bibinfo {author}
  {\bibfnamefont {F.}~\bibnamefont {Pedregosa}}, \bibinfo {author}
  {\bibfnamefont {P.}~\bibnamefont {{van Mulbregt}}},\ and\ \bibinfo {author}
  {\bibnamefont {{SciPy 1.0 Contributors}}},\ }\href
  {https://doi.org/10.1038/s41592-019-0686-2} {\bibfield  {journal} {\bibinfo
  {journal} {Nature Methods}\ }\textbf {\bibinfo {volume} {17}},\ \bibinfo
  {pages} {261} (\bibinfo {year} {2020})}\BibitemShut {NoStop}%
\bibitem [{\citenamefont {Reback}\ \emph {et~al.}(2020)\citenamefont {Reback},
  \citenamefont {McKinney}, \citenamefont {jbrockmendel}, \citenamefont {den
  Bossche}, \citenamefont {Augspurger}, \citenamefont {Cloud}, \citenamefont
  {gfyoung}, \citenamefont {Sinhrks}, \citenamefont {Klein}, \citenamefont
  {Roeschke}, \citenamefont {Hawkins}, \citenamefont {Tratner}, \citenamefont
  {She}, \citenamefont {Ayd}, \citenamefont {Petersen}, \citenamefont {Garcia},
  \citenamefont {Schendel}, \citenamefont {Hayden}, \citenamefont
  {MomIsBestFriend}, \citenamefont {Jancauskas}, \citenamefont {Battiston},
  \citenamefont {Seabold}, \citenamefont {chris b1}, \citenamefont
  {h~vetinari}, \citenamefont {Hoyer}, \citenamefont {Overmeire}, \citenamefont
  {alimcmaster1}, \citenamefont {Dong}, \citenamefont {Whelan},\ and\
  \citenamefont {Mehyar}}]{jeff_reback_2020_3715232}%
  \BibitemOpen
  \bibfield  {author} {\bibinfo {author} {\bibfnamefont {J.}~\bibnamefont
  {Reback}}, \bibinfo {author} {\bibfnamefont {W.}~\bibnamefont {McKinney}},
  \bibinfo {author} {\bibnamefont {jbrockmendel}}, \bibinfo {author}
  {\bibfnamefont {J.~V.}\ \bibnamefont {den Bossche}}, \bibinfo {author}
  {\bibfnamefont {T.}~\bibnamefont {Augspurger}}, \bibinfo {author}
  {\bibfnamefont {P.}~\bibnamefont {Cloud}}, \bibinfo {author} {\bibnamefont
  {gfyoung}}, \bibinfo {author} {\bibnamefont {Sinhrks}}, \bibinfo {author}
  {\bibfnamefont {A.}~\bibnamefont {Klein}}, \bibinfo {author} {\bibfnamefont
  {M.}~\bibnamefont {Roeschke}}, \bibinfo {author} {\bibfnamefont
  {S.}~\bibnamefont {Hawkins}}, \bibinfo {author} {\bibfnamefont
  {J.}~\bibnamefont {Tratner}}, \bibinfo {author} {\bibfnamefont
  {C.}~\bibnamefont {She}}, \bibinfo {author} {\bibfnamefont {W.}~\bibnamefont
  {Ayd}}, \bibinfo {author} {\bibfnamefont {T.}~\bibnamefont {Petersen}},
  \bibinfo {author} {\bibfnamefont {M.}~\bibnamefont {Garcia}}, \bibinfo
  {author} {\bibfnamefont {J.}~\bibnamefont {Schendel}}, \bibinfo {author}
  {\bibfnamefont {A.}~\bibnamefont {Hayden}}, \bibinfo {author} {\bibnamefont
  {MomIsBestFriend}}, \bibinfo {author} {\bibfnamefont {V.}~\bibnamefont
  {Jancauskas}}, \bibinfo {author} {\bibfnamefont {P.}~\bibnamefont
  {Battiston}}, \bibinfo {author} {\bibfnamefont {S.}~\bibnamefont {Seabold}},
  \bibinfo {author} {\bibnamefont {chris b1}}, \bibinfo {author} {\bibnamefont
  {h~vetinari}}, \bibinfo {author} {\bibfnamefont {S.}~\bibnamefont {Hoyer}},
  \bibinfo {author} {\bibfnamefont {W.}~\bibnamefont {Overmeire}}, \bibinfo
  {author} {\bibnamefont {alimcmaster1}}, \bibinfo {author} {\bibfnamefont
  {K.}~\bibnamefont {Dong}}, \bibinfo {author} {\bibfnamefont {C.}~\bibnamefont
  {Whelan}},\ and\ \bibinfo {author} {\bibfnamefont {M.}~\bibnamefont
  {Mehyar}},\ }\href {https://doi.org/10.5281/zenodo.3715232} {\bibinfo {title}
  {pandas-dev/pandas: Pandas 1.0.3}} (\bibinfo {year} {2020})\BibitemShut
  {NoStop}%
\bibitem [{\citenamefont {{W}es
  {M}c{K}inney}(2010)}]{mckinney-proc-scipy-2010}%
  \BibitemOpen
  \bibfield  {author} {\bibinfo {author} {\bibnamefont {{W}es {M}c{K}inney}},\
  }in\ \href {https://doi.org/10.25080/Majora-92bf1922-00a} {\emph {\bibinfo
  {booktitle} {{P}roceedings of the 9th {P}ython in {S}cience {C}onference}}},\
  \bibinfo {editor} {edited by\ \bibinfo {editor} {\bibnamefont {{S}t\'efan
  van~der {W}alt}}\ and\ \bibinfo {editor} {\bibnamefont {{J}arrod
  {M}illman}}}\ (\bibinfo {year} {2010})\ pp.\ \bibinfo {pages} {56 --
  61}\BibitemShut {NoStop}%
\bibitem [{\citenamefont {Lepage}(2020)}]{peter_lepage_2020_4290884}%
  \BibitemOpen
  \bibfield  {author} {\bibinfo {author} {\bibfnamefont {G.~P.}\ \bibnamefont
  {Lepage}}\ }\href {https://doi.org/doi:10.5281/zenodo.4290884}
  {doi:10.5281/zenodo.4290884} (\bibinfo {year} {2020}),\ \Eprint
  {https://arxiv.org/abs/https://github.com/gplepage/gvar}
  {https://github.com/gplepage/gvar} \BibitemShut {NoStop}%
\bibitem [{\citenamefont {{Wolfram Research Inc.}}()}]{Mathematica}%
  \BibitemOpen
  \bibfield  {author} {\bibinfo {author} {\bibnamefont {{Wolfram Research
  Inc.}}},\ }\href@noop {} {\bibinfo {title} {Mathematica, {V}ersion 14.0}},\
  \bibinfo {note} {{\tt https://www.wolfram.com/mathematica}}\BibitemShut
  {NoStop}%
\bibitem [{\citenamefont {Hunter}(2007)}]{Hunter:2007}%
  \BibitemOpen
  \bibfield  {author} {\bibinfo {author} {\bibfnamefont {J.~D.}\ \bibnamefont
  {Hunter}},\ }\href {https://doi.org/10.1109/MCSE.2007.55} {\bibfield
  {journal} {\bibinfo  {journal} {Computing in Science \& Engineering}\
  }\textbf {\bibinfo {volume} {9}},\ \bibinfo {pages} {90} (\bibinfo {year}
  {2007})}\BibitemShut {NoStop}%
\end{thebibliography}%


\end{document}